\documentclass[apj]{emulateapj}

\usepackage{apjfonts}
\usepackage{epsfig}
\usepackage{graphicx}
\usepackage{url}
\usepackage{natbib}
\usepackage{float}
\usepackage{amsmath}
\usepackage[bf, sf, FIGTOPCAP, nooneline, tight]{subfigure}
\usepackage{graphicx}
\usepackage{dcolumn}
\usepackage{bm}
\usepackage[usenames,dvipsnames]{color}
\usepackage[citebordercolor={0 1 0}]{hyperref}
\usepackage{amssymb}
\newcommand{\be}{\begin{equation}}
\newcommand{\ee}{\end{equation}}
\newcommand{\ba}{\begin{eqnarray}}
\newcommand{\ea}{\end{eqnarray}}

\def\({\left(}
\def\){\right)}

\begin{document}

\shorttitle{Investigation on the slowing down of cosmic acceleration}
\shortauthors{Wang, S. et al.}

\title{A comprehensive investigation on the slowing down of cosmic acceleration}

\author{
Shuang Wang\altaffilmark{1},
Yazhou Hu\altaffilmark{1,2,3},
Miao Li\altaffilmark{1},
Nan Li\altaffilmark{1,2,3},
}

\email{wangshuang@mail.sysu.edu.cn (Corresponding author)}
\email{asiahu@itp.ac.cn}
\email{limiao9@mail.sysu.edu.cn}
\email{linan@itp.ac.cn}
\altaffiltext{1}{School of Astronomy and Space Science, Sun Yat-Sen University, Guangzhou 510275, P. R. China}
\altaffiltext{2}{Institute of Theoretical Physics, Chinese Academy of Sciences,
Beijing 100190, P. R. China}
\altaffiltext{3}{Kavli Institute of Theoretical Physics China, Chinese Academy of Sciences, Beijing 100190, P. R. China}

\begin{abstract}

Shafieloo ea al. firstly proposed the possibility that the current cosmic acceleration (CA) is slowing down. However, this is rather counterintuitive because a slowing down CA cannot be accommodated in most mainstream cosmological models. In this work, by exploring the evolutionary trajectories of dark energy equation of state $w(z)$ and deceleration parameter $q(z)$, we present a comprehensive investigation on the slowing down of CA from both the theoretical and the observational sides. For the theoretical side, we study the impact of different $w(z)$ by using six parametrization models, and then discuss the effects of spatial curvature. For the observational side, we investigate the effects of different type Ia supernovae (SNe Ia), different baryon acoustic oscillation (BAO), and different cosmic microwave background (CMB) data, respectively.
We find that
(1) The evolution of CA are insensitive to the specific form of $w(z)$; in contrast, a non-flat Universe more favors a slowing down CA than a flat Universe.
(2) SNLS3 SNe Ia datasets favor a slowing down CA at 1$\sigma$ confidence level, while JLA SNe Ia samples prefer an eternal CA; in contrast,
the effects of different BAO data are negligible. 
(3) Compared with CMB distance prior data, full CMB data more favor a slowing down CA.
(4) Due to the low significance, the slowing down of CA is still a theoretical possibility that cannot be confirmed by the current observations.

\end{abstract}


\section{Introduction}

Since its discovery in 1998~\citep{Riess1998, Perl1999}, cosmic acceleration (CA) has been confirmed by various cosmological observations,
such as type Ia supernovae (SNe Ia)~\citep{Hicken2009, Conley2011, Suzuki2012, Betoule2014},
baryon acoustic oscillation (BAO)~\citep{Tegmark2004, Eisenstein2005, Percival2010, Padmanabhan2012},
cosmic microwave background (CMB)~\citep{Spergel2003, Spergel2007, Hinshaw2013, Planck201513}, and so on.
Now CA has become one of the central problems
in modern cosmology~\citep{Caldwell2009, Wang2010, LiLiWangWang2011, Bamba2012, LiLiWangWang2013, Weinberg2013}.
Currently, almost all the mainstream cosmological models,
such as $\Lambda$-cold-dark-matter ($\Lambda$CDM) model,
holographic dark energy model~\citep{Li2004, HuangLi2004, LiLiWang2009a, LiLiWang2009b, Zhang2012, LiYH2013},
and various scalar field dark energy (DE) models~\citep{Zlatev1999, Armendariz-Picon1999, Kamenshchik2001, Caldwell2002, Padmanabhan2002, WZ2008, WZX2008}, predict an eternal CA.

In~\citep{Shafieloo2009}, Shafieloo, Sanhi, and Starobinsky firstly proposed the possibility that the current CA is slowing down.
By analysing the Constitution SNe Ia sample~\citep{Hicken2009} with the Chevallier-Polarski-Linder (CPL) model~\citep{Chevallier2001, Linder2003},
they found that the CA has already peaked and is now slowing down at 1$\sigma$ confidence level (CL).
This result is quite interesting,
because a slowing down CA cannot be accommodated in almost all the mainstream models,
and only some strange doomsday models~\citep{YunWang2004} can explain this extremely counterintuitive phenomenon.

In recent years, this topic has attracted
a lot of interests~\citep{Huang2009, Gong2010, LiWuYu2011, Lin2013, Cardenas2012, Cardenas2013, Cardenas2014}
For example, in~\citep{LiWuYu2011}, using the CPL model,
the authors found that the Union2 SNe Ia data~\citep{Amanullah2010} favor a slowing down CA at 1$\sigma$ CL.
In~\citep{Cardenas2012}, also using the CPL model,
the authors found that the combined Union2+BAO+CMB data also prefer a slowing down CA if the spatial curvature is taken into account.
In a recent paper, in the framework of the CPL model,
the Planck Collaboration~\citep{Planck201514} reconstructed the evolution of DE equation of state (EoS) $w(z)$
by using a combination of Planck, galaxy weak lensing and redshift space distortions data;
they found that $w(z)$ is a decreasing function of redshift $z$ when $z \rightarrow 0$ (See Fig. 5 of ~\citep{Planck201514}),
which also indicates a slowing down CA.

Although the slowing down of CA has been widely investigated,
previous studies mainly focus on the effects of different SNe Ia datasets;
in addition, most of these works only consider the CPL model in a flat Universe.
A relatively comprehensive study was given in~\citep{Magana2014},
where Magana et al. investigated the evolutionary trajectories of deceleration parameter $q(z)$
by using five DE parametrization models and four SNe Ia datasets.
In addition, they also studied the impact of adding CMB and BAO data on the evolution of $q(z)$.
However, some other factors are not considered in Ref.~\citep{Magana2014}.
For example, the impacts of spatial curvature, which may have a degeneracy with the DE EoS $w(z)$~\citep{Clarkson2007}, are not taken into account.
In addition, the effects of different BAO and CMB data on the evolutionary behavior of CA are not studied, either.
To make a comprehensive and systematic investigation on this topic, all the factors mentioned above need to be taken into account,
and this is the main aim of our work.

For the theoretical side, we study the impact of different $w(z)$ by using six popular DE parametrization models,
including the CPL model, the Jassal-Bagla-Padmanabhan (JBP) model~\citep{Jassal2005a, Jassal2005b}, the Barbosa-Alcaniz (BA) model~\citep{Barboza2008}, the Ma-Zhang (MZ) model~\citep{MaZhang2011}, the Feng-Shen-Li-Li (FSLL) model~\citep{FSLL2012} and the Wang (WANG) model~\citep{YunWang2008}.
In addition, we also discuss the effects of spatial curvature on this topic.
For the observational side,
we investigate the effects of three kinds of SNe Ia data, two kinds of BAO data, and four kinds of CMB data, respectively.
The detailed information of these observational data are listed in table~\ref{tab:dataset}.

\begin{table*}
\centering
\caption{All the observational data used in the present work}
\label{tab:dataset}
\centering
\begin{tabular}{p{3.5cm}|c|cccccccccc}

\hline\hline
Observation         ~&~       Data                      & Reference    \\
  \hline
  &&\\
  SNe Ia        ~&~SNLS3 (constant $\beta$) $\ ^a$        &~\cite{Conley2011} \\
  &&\\
                ~&~SNLS3 (linear $\beta$) $\ ^b$  &~\cite{Conley2011,WangWang2013}\\
  &&\\
                       ~&~JLA                  &~\cite{Betoule2014}\\
  \hline
  &&\\
  BAO             ~&~6dfGS(1D) $\ ^c$                        &~\cite{Beutler2011} \\
  &&\\
             ~&~SDSS DR7 (1D) $\ ^c$        &~\cite{Padmanabhan2012}\\
  &&\\
             ~&~BOSS DR9 (1D) $\ ^c$        &~\cite{Eisenstein2011} \\
  &&\\
             ~&~SDSS DR7 (2D) $\ ^d$        &~\cite{Hemantha2014}\\
  &&\\
             ~&~BOSS DR9 (2D) $\ ^d$        &~\cite{Wangyun2014} \\
  &&\\

  \hline
  &&\\
  CMB        ~&~Planck 2015 distance priors &~\cite{Planck201514} \\
  &&\\
             ~&~Planck 2013 distance priors &~\cite{WangyunWangshuang2013} \\
  &&\\
             ~&~WMAP9 distance priors       & ~\cite{WangyunWangshuang2013}\\
  &&\\
             ~&~Planck 2015 full data       &~\cite{Planck201511}\\

  \hline\hline
\end{tabular}
\leftline{\noindent$\ ^a$ ``constant $\beta$'' means that a constant supernova color-luminosity parameter $\beta$ is used in the analysis.}
\leftline{\noindent$\ ^b$ ``linear $\beta$'' means that a time-varying supernova color-luminosity parameter $\beta$ is used in the analysis.}
\leftline{\noindent$\ ^c$ ``1D'' means that the BAO data are obtained by using the spherically averaged one-dimensional (1D) galaxy clustering statistics.}
\leftline{\noindent$\ ^d$ ``2D'' means that the BAO data are obtained by using the anisotropic two-dimensional (2D) galaxy clustering statistics.}
\end{table*}

We present our method in Section~\ref{sec:method}, our results in Section~\ref{sec:results}, and
summarize and conclude in Section~\ref{sec:conclusion}.

\section{Methodology}
\label{sec:method}

In this section, we briefly review the theoretical framework of various DE models, and describe the observational data used in the present work.

\subsection{Theoretical Models}
\label{subsec:theoretical models}

In a non-flat Universe, the Friedmann equation is
\be\label{F.e.}
    3M_{pl}^{2}H^{2}=\rho_{r}+\rho_{m}+\rho_{k}+\rho_{de},
\ee
where $H \equiv \dot{a}/a$ is the Hubble parameter,
$a=(1+z)^{-1}$ is the scale factor of the Universe (we take today's scale factor $a_0=1$),
the dot denotes the derivative with respect to cosmic time $t$,
$M^2_{pl} = (8\pi G)^{-1}$ is the reduced Planck mass, $G$ is Newtonian gravitational constant,
$\rho_{r}$, $\rho_{m}$, $\rho_{k}$ and $\rho_{de}$
are the energy densities of radiation, matter, spatial curvature and DE, respectively.
The reduced Hubble parameter $E(z)\equiv H(z)/H_{0}$ satisfies
\be\label{eq:E}
E^{2}=\Omega_{r0}(1+z)^{4}+\Omega_{m0}(1+z)^{3}+\Omega_{k0}(1+z)^2+\Omega_{de0}f(z),
\ee
where $H_0 = 100h ~ km \cdot s^{-1} \cdot Mpc^{-1}$ is the Hubble constant,
$h$ is the dimensionless Hubble constant, $\Omega_{r0}$, $\Omega_{m0}$, $\Omega_{k0}$ and $\Omega_{de0}$
are the present fractional densities of radiation, matter, spatial curvature and DE, respectively.
Per \citep{WangWang2013}, we take $\Omega_{r0}=\Omega_{m0} / (1+z_{\rm eq})$, where $z_{\rm eq}=2.5\times 10^4 \Omega_{m0} h^2 (T_{\rm cmb}/2.7\,{\rm K})^{-4}$ and $T_{\rm cmb}=2.7255\,{\rm K}$.
Since $\Omega_{de0}=1-\Omega_{m0}-\Omega_{r0}-\Omega_{k0}$, $\Omega_{de0}$ is not an independent parameter.
Here the DE density function $f(z) \equiv \rho_{de}(z)/\rho_{de}(0)$, which satisfies
\be\label{Density function.e.}
    f(z)={\rm exp}\Big[3\int_{0}^{z}dz^{\prime}\frac{1+w(z^{\prime})}{1+z^{\prime}}\Big],
\ee
where the EoS $w$ is the ratio of pressure to density for the DE
\be
\label{eq:eos}
w = p_{de}/\rho_{de}.
\ee

To study the impacts of different $w(z)$, here we consider six popular parametrization models:
\footnote{In addition to assuming a specific parametrization form for $w(z)$,
another popular method of exploring the dynamical evolution of EoS is adopting a binned parametrization\cite{Huterer03,Huterer05,WLL10,WLL11,Li11,Gong13}.
For simplicity, here we do not study the case of binned parametrization.}
\begin{itemize}
\item
CPL model \citep{Chevallier2001, Linder2003} has a dynamical EoS $w(z) = w_{0} + w_{a}\frac{z}{1+z}$,
thus we have
\ba\label{E_cpl}
E(z)=\Big(\Omega_{r0}(1+z)^{4}+\Omega_{m0}(1+z)^{3}+\Omega_{k0}(1+z)^2
\nonumber\\
+\Omega_{de0}(1+z)^{3(1+w_0+w_a)}e^{\frac{-3w_az}{1+z}}\Big)^{1/2},
\ea
\item
JBP model~\citep{Jassal2005a, Jassal2005b} has a dynamical EoS $w(z) = w_{0} + w_{a}\frac{z}{(1+z)^2}$,
thus we have
\ba\label{E_jbp}
E(z)=\Big(\Omega_{r0}(1+z)^{4}+\Omega_{m0}(1+z)^{3}+\Omega_{k0}(1+z)^2
\nonumber\\
+\Omega_{de0}(1+z)^{3(1+w_0)}e^{\frac{3w_az^2}{2(1+z)^2}}\Big)^{1/2},
\ea
\item
BA model~\citep{Barboza2008} has a dynamical EoS $w(z) = w_{0} + w_{a}\frac{z(1+z)}{1+z^2}$,
thus we have
\ba\label{E_ba}
E(z)=\Big(\Omega_{r0}(1+z)^{4}+\Omega_{m0}(1+z)^{3}+\Omega_{k0}(1+z)^2
\nonumber\\
+\Omega_{de0}(1+z)^{3(1+w_0)}(1+z^2)^{\frac{3w_a}{2}}\Big)^{1/2},
\ea
\item
MZ model~\citep{MaZhang2011} has a dynamical EoS $w(z) = w_{0} + w_{a}(\frac{ln(2+z)}{1+z}-ln2)$,
thus we have
\ba\label{E_mz}
E(z)=\Big(\Omega_{r0}(1+z)^{4}+\Omega_{m0}(1+z)^{3}+\Omega_{k0}(1+z)^2
\nonumber\\
+\Omega_{de0}4^{3w_a}(1+z)^{3(1+w_0+w_a-w_aln2)}(2+z)^{\frac{-3w_a(2+z)}{1+z}}\Big)^{1/2},
\ea
\item
FSLL model~\citep{FSLL2012} has a dynamical EoS $w(z) = w_{0} + w_{a}\frac{z}{1+z^2}$,
thus we have
\ba\label{E_fsll}
E(z)=\Big(\Omega_{r0}(1+z)^{4}+\Omega_{m0}(1+z)^{3}+\Omega_{k0}(1+z)^2
\nonumber\\
+\Omega_{de0}(1+z)^{3(1+w_0-0.5w_a)}(1+z^2)^{0.75w_a}e^{1.5w_aarctanz}\Big)^{1/2},
\ea
\item
WANG model~\citep{YunWang2008} has a dynamical EoS $w(z) = w_{0}\frac{1-2z}{1+z} + w_{a}\frac{z}{(1+z)^2}$,
thus we have
\ba\label{E_wang}
E(z)=\Big(\Omega_{r0}(1+z)^{4}+\Omega_{m0}(1+z)^{3}+\Omega_{k0}(1+z)^2
\nonumber\\
+\Omega_{de0}(1+z)^{3(1-2w_0+3w_a)}e^{\frac{9(w_0-w_a)z}{1+z}}\Big)^{1/2}.
\ea
\end{itemize}
For each model, the expression of $E(z)$ will be used to calculate the observational quantities appearing in the next subsection.
To make a comparison, the simplest $\Lambda$CDM model and $w$CDM model are also adopted in the analysis.
In addition, we also use the deceleration parameter $q\equiv -\frac{\ddot{a}}{aH^{2}}$ to investigate the evolutionary behavior of CA.

\subsection{Observational Data}
\label{subsec:observational data}

In this subsection, we describe how to include various observational data into the $\chi^2$ statistics.

\subsubsection{SNe Ia Data}

Here we use the SNLS3 ``Combined'' samples (consisting of 472 SNe Ia)~\citep{Conley2011} and the JLA samples (consisting of 740 SNe Ia)~\citep{Betoule2014}.

The $\chi^2$ function of the SNLS3 supernova (SN) data is given by~\citep{Conley2011}
\begin{equation}\label{SNchisq}
\chi^2_{SNLS3}=\Delta \overrightarrow{\bf m}^T \cdot {\bf C}^{-1} \cdot \Delta \overrightarrow{\bf m},
\end{equation}
where $\Delta {\overrightarrow {\bf m}} = {\overrightarrow {\bf m}}_B - {\overrightarrow {\bf m}}_{\rm mod}$ is a vector of
model residuals of the SN sample,
and $m_B$ is the rest-frame peak $B$ band magnitude of the SN.

Without considering the possibility of potential SN evolution,
the predicted magnitude of SN can be expressed as
\be~\label{snvec_betaz}
m_{\rm mod}=5 \log_{10}{\cal D}_L(z) - \alpha (s-1) +\beta {\cal C} + {\cal M},
\ee
where $\alpha$ and $\beta$ are SN stretch-luminosity parameter and SN color-luminosity parameter,
$s$ and ${\cal C}$ are stretch measure and color measure for the SN light curve,
$\mathcal{M}$ is a parameter representing some combination of the absolute magnitude $M$ of a fiducial SNe Ia and the Hubble constant $H_0$.
\footnote{It must be emphasized that, in order to include host-galaxy information in the cosmological fits,
Conley et al. \citep{Conley2011} split the SNLS3 sample based on host-galaxy stellar mass at $10^{10} M_{\odot}$,
and made ${\cal M}$ to be different for the two samples.
So there are two values of ${\cal M}$ (i.e. ${\cal M}_1$ and ${\cal M}_2$) for the SNLS3 data.
Moreover, Conley et al. removed ${\cal M}_1$ and ${\cal M}_2$ from cosmology-fits by analytically marginalizing over them
(for more details, see the appendix C of~\citep{Conley2011}).
In the present work, we just follow the recipe of~\citep{Conley2011}, and do not treat ${\cal M}$ as model parameter.}

It must be emphasized that,
the current studies on various SNe Ia data sets,
including SNLS3~\citep{WangWang2013}, Union2.1~\citep{Mohlabeng2013}, and Pan-STARRS1~\citep{Scolnic2014},
all indicated that, although $\alpha$ is still consistent with a constant,
$\beta$ should evolve along with redshift $z$ at very high confidence level (CL).
Moreover, the evolution of $\beta$ has significant effects on parameter estimation of various cosmological models~\citep{WLZ2014,WWGZ2014,WWZ2014,Wang2015,LLWZ2016}.
In addition, it has been proved that~\citep{WangWang2013} the fitting results are insensitive to the functional form of $\beta(z)$ assumed.
So per~\citep{WLZ2014}, we also consider the case of a constant $\alpha$ and a linear $\beta = \beta_0 + \beta_1 z$,
then the predicted magnitude of SN becomes
\be~\label{snvec_betaz}
m_{\rm mod}=5 \log_{10}{\cal D}_L(z) - \alpha (s-1) +\beta(z) {\cal C} + {\cal M}.
\ee

So there are two ways to deal with the SNLS3 data: adopting a constant $\beta=\beta_{0}$ and adopting a linear $\beta = \beta_0 + \beta_1 z$.
For simplicity, hereafter we will call them ``SNLS3 (constant $\beta$)'' and ``SNLS3 (linear $\beta$)'' data, respectively.

The luminosity distance ${\cal D}_L(z)$ is defined as
\be~\label{eq:sn_lumdist}
{\cal D}_L(z)\equiv H_0c^{-1} (1+z_{\rm hel}) r(z),
\ee
where $c$ is the speed of light,
$z$ and $z_{\rm hel}$ are the CMB restframe and heliocentric redshifts of SN,
and the comoving distance $r(z)$ is given by
\be
\label{eq:rz}
r(z)=cH_0^{-1}\, |\Omega_k|^{-1/2} {\rm sinn}[|\Omega_k|^{1/2}\, \Gamma(z)].
\ee
Here $\Gamma(z)=\int_0^z\frac{dz'}{E(z')}$,
${\rm sinn}(x)=\sin(x)$, $x$, $\sinh(x)$ for $\Omega_k<0$, $\Omega_k=0$, and $\Omega_k>0$, respectively.

The total covariance matrix $\mbox{\bf C}$, which appears in Eq.~\ref{SNchisq}, can be written as~\citep{Conley2011}
\be
\mbox{\bf C}=\mbox{\bf D}_{\rm stat}+\mbox{\bf C}_{\rm stat}+\mbox{\bf C}_{\rm sys}.
\ee
Here $\mbox{\bf D}_{\rm stat}$ denotes the diagonal part of the statistical uncertainty,
$\mbox{\bf C}_{\rm stat}$ and $\mbox{\bf C}_{\rm sys}$ denote the statistical and systematic covariance matrices, respectively.
Notice that the covariance matrices of the ``SNLS3 (constant $\beta$)'' and the ``SNLS3 (linear $\beta$)'' data are different.
For the details of constructing the total covariance matrix $\mbox{\bf C}$, see \cite{Conley2011}.

In addition, we also use the JLA SN samples~\citep{Betoule2014}.
Since the $\chi^2$ function of the JLA samples has the similar form with the case of the ``SNLS3 (constant $\beta$)'' data,
for simplicity we do not describe how to include the JLA data into the $\chi^2$ statistics any more.
More details about the JLA samples can be found in~\citep{Betoule2014}.

\subsubsection{BAO Data}

Here we use two types of BAO data that are extracted by using the spherically averaged 1D galaxy clustering statistics
and the anisotropic 2D galaxy clustering statistics, respectively.
For simplicity, hereafter we will call them ``BAO(1D)'' and ``BAO(2D)'' data, respectively.

For the BAO observation, two characteristic quantities,
$D_V(z)$ and $r_s(z_d)$, are often adopted to constrain various cosmological models.
The volume-averaged effective distance $D_V(z)$ is given by ~\citep{Eisenstein2005},
\be
D_V(z) \equiv [(1+z)^2{D_A(z)}^2\frac{cz}{H(z)}]^{1/3},
\ee
where the angular diameter distance
\be
D_{A}(z) = c H_0^{-1}r(z)/(1+z),
\ee
and the comoving distance $r(z)$ is given by Eq.~\ref{eq:rz}.
In addition, the comoving sound horizon $r_s(z)$ is given by \citep{WangWang2013}
\be~\label{sound_horizon}
r_s(z) = cH_0^{-1}\int_{0}^{a}\frac{da^{\prime}}{\sqrt{3(1+\overline{R_b}a^\prime){a^\prime}^4E^2(z^\prime)}},
\ee
where $\overline{R_b}=31500\Omega_{b0}h^2(T_{CMB}/2.7K)^{-4}$, and $\Omega_{b0}$ is the present fractional density of baryon.
The redshift of the drag epoch $z_d$ is given in Ref. \citep{EisensteinHu1998}.

The BAO(1D) data include three data points that are extracted from the 6dFGS (6dF)~\citep{Beutler2011},
the Sloan Digital Sky Survey Data Release 7 (SDSS DR7)~\citep{Abazajian2009}
and the Baryon Oscillation Spectroscopic Survey Data Release 9 (BOSS DR9)~\citep{Eisenstein2011}, respectively.
Making use of the data of 6df, Beutler {et~al.}~\citep{Beutler2011} gave
\be
r_s(z_d)/D_V(0.106)^{data} = 0.336\pm0.015.
\ee
Adopting the data of SDSS DR7,
Padmanabhan {et~al.} ~\citep{Padmanabhan2012} got
\be
D_V(0.35)/r_s(z_d)^{data} = 8.88\pm0.17.
\ee
Based on the data of BOSS DR9,
Anderson {et~al.} \citep{Anderson2012} obtained
\be
D_V(0.57)/r_s(z_d)^{data} = 13.67\pm0.22.
\ee
The $\chi^2$ function of the BAO(1D) data can be expressed as \footnote{Since the redshifts of these three surveys are separated widely,
it is reasonable to neglect all the correlations among these surveys.}
\be
\label{eq:chi2bao1}
\chi^2_{BAO(1D)}=\sum_{i=1}^{3}\Big(\frac{q_i - q_i^{data}}{\sigma_i}\Big)^2.
\ee
Here $q_1=r_{\rm s}(z_d)/D_{\rm V}(0.106)$, $q_2=D_{\rm V}(0.35)/r_{\rm s}(z_d)$, and $q_3=D_{\rm V}(0.57)/r_{\rm s}(z_d)$.
In addition, $\sigma_i$ denotes the standard deviation of data, and $i=1,2,3$.

For the BAO(2D) data, we use two observed quantities, $H(z)r_s(z_d)/c$ and $D_A(z)/r_s(z_d)$, that are extracted from SDSS DR7 and BOSS DR9 data, respectively.
Making use of the 2D matter power spectrum of SDSS DR7 samples,
Hemantha, Wang, and Chuang~\citep{Hemantha2014} got
\ba
H(0.35)r_s(z_d)/c^{data}&=&0.0431  \pm  0.0018,  \nonumber \\
D_A(0.35)/r_s(z_d)^{data}&=& 6.48  \pm  0.25.
\label{eq:CW2}
\ea
In addition, using the 2D two point correlation functions of BOSS DR9 samples,
Wang~\cite{Wangyun2014} obtained
\ba
H(0.57)r_s(z_d)/c^{data}&=&0.0444	\pm  0.0019,  \nonumber \\
D_A(0.57)/r_s(z_d)^{data}&=& 9.01	\pm  0.23.
\label{eq:C13}
\ea
The $\chi^2$ function of the BAO(2D) data is
\be
\chi^2_{BAO(2D)}=\chi^2_{1}+\chi^2_{2},
\ee
with $z_{1}=0.35$ and $z_{2}=0.57$.
Here
\be
\label{eq:chi2bao2}
\chi^2_{i}=\Delta Q_i \left[ {\rm C}^{-1}_{i}(Q_i,Q_j)\right] \Delta Q_j,
\hskip .2cm
\Delta Q_i= Q_i - Q_i^{data},
\ee
where $Q_1=H(z_{i})r_s(z_d)/c$, $Q_2=D_A(z_{i})/r_s(z_d)$, and $i=1,2$.
Based on Refs.~\citep{Hemantha2014} and~\citep{Wangyun2014}, we get
\begin{equation}
 {\rm C}_{1}=\left(
  \begin{array}{cc}
    0.00000324 & -0.00010728 \\
    -0.00010728 & 0.0625 \\
  \end{array}
\right),
\end{equation}
\begin{equation}
 {\rm C}_{2}=\left(
  \begin{array}{cc}
    0.00000361 & 0.0000176111 \\
    0.0000176111 & 0.0529 \\
  \end{array}
\right).
\end{equation}

\subsubsection{CMB Data}

For CMB data, we use the distance priors data extracted from Planck 2015~\citep{Planck201511, Planck201513, Planck201514},
Planck 2013 \citep{WangyunWangshuang2013} and WMAP9 \citep{WangyunWangshuang2013} samples.
To make a comparison, we also use the full ``$Planck TT+lowP$'' data given by the Planck 2015 data release~\citep{Planck201511}.
For simplicity, hereafter we will call them ``Planck2015'', ``Planck2013'', ``WMAP9'' and ``Planck2015(full data)'', respectively.

CMB gives us the comoving distance to the photon-decoupling surface $r(z_*)$ and the comoving sound horizon at photon-decoupling epoch $r_s(z_*)$.
The redshift of the photon-decoupling epoch $z_*$ is given in Ref. \citep{HuSugiyama1996}.
Wang and Mukherjee~\citep{WangMukherjee2007} have shown that the CMB shift parameters
\ba
l_a &\equiv &\pi r(z_*)/r_s(z_*), \nonumber\\
R &\equiv &\sqrt{\Omega_{m0} H_0^2} \,r(z_*)/c,
\ea
together with $\omega_b\equiv \Omega_{b0} h^2$, provide an efficient summary of CMB data.
Therefore, the $\chi^2$ function for the CMB distance prior data can be expressed as
\be
\label{eq:chi2CMB}
\chi^2_{CMB}=\Delta p_i \left[ \mbox{Cov}^{-1}_{CMB}(p_i,p_j)\right]
\Delta p_j,
\hskip .2cm
\Delta p_i= p_i - p_i^{data},
\ee
where $p_1=R(z_*)$, $p_2=l_a(z_*)$, and $p_3=\omega_b$.
The covariance matrix for $(p_1, p_2, p_3)$ is given by
\be
\mbox{Cov}_{CMB}(p_i,p_j)=\sigma(p_i)\, \sigma(p_j) \,\mbox{NormCov}_{CMB}(p_i,p_j),
\label{eq:CMB_cov}
\ee
where $\sigma(p_i)$ is the 1$\sigma$ error of observed quantity $p_i$,
$\mbox{NormCov}_{CMB}(p_i,p_j)$ is the corresponding normalized covariance matrix, and $i,j=1,2,3$.

As mentioned above, in the present work we use three kinds of CMB distance prior data:
\begin{itemize}
\item
The Planck 2015 data are~\citep{Planck201514}
\ba
&& p_{1}^{data} = 1.7382\pm0.0088, \nonumber\\
&& p_{2}^{data} = 301.63\pm0.15, \nonumber\\
&& p_{3}^{data} = 0.02262\pm0.00029.
\label{eq:CMB_mean_planck}
\ea
In addition,
\be
\mbox{NormCov}_{Planck 2015}=
\left(
\begin{array}{ccc}
  1.00  &    0.64  &   -0.75    \\
  0.64  &    1.00  &   -0.55    \\
 -0.75  &   -0.55  &    1.00    \\
\end{array}
\right).
\label{eq:normcov_planck}
\ee
\item
The Planck 2013 data are~\citep{WangyunWangshuang2013}
\ba
&& p_{1}^{data} = 301.57\pm0.18, \nonumber\\
&& p_{2}^{data} = 1.7407\pm0.0094, \nonumber\\
&& p_{3}^{data} = 0.02228\pm0.00030.
\label{eq:CMB_mean_planck_2013}
\ea
In addition,
\be
\mbox{NormCov}_{Planck 2013}=
\left(
\begin{array}{ccc}
  1.00  &    0.5250  &   -0.4235    \\
  0.5250  &    1.00  &   -0.6925    \\
 -0.4235  &   -0.6925  &    1.00    \\
\end{array}
\right).
\label{eq:normcov_planck_2013}
\ee
\item
The WMAP9 data are~\citep{WangyunWangshuang2013}
\ba
&& p_{1}^{data} = 302.02\pm0.66, \nonumber\\
&& p_{2}^{data} = 1.7327\pm0.0164, \nonumber\\
&& p_{3}^{data} = 0.02260\pm0.00053.
\label{eq:CMB_mean_wmap9}
\ea
In addition,
\be
\mbox{NormCov}_{WMAP9}=
\left(
\begin{array}{ccc}
  1.00  &    0.3883  &   -0.6089    \\
  0.3883  &    1.00  &   -0.5239    \\
 -0.6089  &   -0.5239  &    1.00    \\
\end{array}
\right).
\label{eq:normcov_wmap9}
\ee
\end{itemize}

To make a comparison, we also use the full ``$Planck TT+lowP$'' data given by the Planck 2015 data release~\citep{Planck201511}
The complete likelihood description of ``Planck2015(full data)'' can be found in Ref.~\citep{Planck201511},
and the corresponding likelihood softwares can be downloaded at http://pla.esac.esa.int/pla/\#results.

\subsubsection{Total $\chi^2$ function}

\begin{table*}
\centering
\caption{the prior ranges of all the free parameters. These prior ranges are uniform for all the models considered in this work.}
\label{tab:priors}
\centering
\begin{tabular}{ccccccccccc}
\hline\hline 
&&\\
Parameter  & $\alpha$ & $\beta_0$ & $\beta_1$ & $\Omega_{k0}$ & $\Omega_{b0}$ & $\Omega_{m0}$ & $h$ & $w_0$ & $w_a$ \\
&&\\ \hline
&&\\
Prior Range                           & $[0,5]$
                                & $[0,5]$
                                & $[0,20]$
                                & $[-1,1]$
                                & $[0,1]$
                                & $[0,1]$
                                & $[0,5]$
                                & $[-6,5]$
                                & $[-15,15]$ $\ ^a$ \\ 
&&\\
\hline
\end{tabular}
\leftline{\noindent$\ ^a$ For the MZ model, the prior range of $w_a$ is $[-35,35]$.}
\end{table*}

The total $\chi^2$ function is
\be
\chi^2=\chi^2_{SN}+\chi^2_{BAO}+\chi^2_{CMB}.
\ee
In the present work, we mainly make use of the ``$\rm SNLS3(linear\ \beta)+BAO(1D)+Planck2015$'' data to study the effects of various theoretical factors on CA.
In addition, other observational data listed in table~\ref{tab:dataset} are also used for comparison.
Finally, we perform an MCMC likelihood analysis using the ``CosmoMC'' package~\citep{Lewis2002}.
In the MCMC analysis, the prior range of each free parameter is uniform for all the DE models. These prior ranges are listed in Table~\ref{tab:priors}.

\section{Results}
\label{sec:results}

In this section, by exploring the evolutionary trajectories of $w(z)$ and $q(z)$,
we study the slowing down of CA from both the theoretical and the observational sides.

\subsection{Effects of Different Theoretical Models on CA}
\label{subsec:de parametrizations}

Here we discuss the effects of different $w(z)$ and spatial curvature on the evolutionary behavior of CA.
To study the impacts of these theoretical factors, it is necessary to fix the observational data used in the analysis first.
As mentioned above, in this subsection we always make use of the ``$\rm SNLS3(linear\ \beta)+BAO(1D)+Planck2015$'' data.

\subsubsection{Impacts of Different $w(z)$}
\label{subsubsec:de parametrizations}

\begin{table*}
\centering
\caption{Fitting results of the $\Lambda$CDM, the $w$CDM and the six DE parametrization models.
Both the best-fit values, the $1\sigma$ errors of various parameters and the $\chi^2_{min}$ of various models are listed.}
\label{tab:res_betaz_cmb1_bao1}
\centering
\begin{tabular}{ccccccccccc}
\hline\hline 
Parameter  & $\Lambda$CDM & $w$CDM & CPL & JBP & BA & MZ & FSLL & WANG \\ \hline
$\alpha$           & $1.434^{+0.098}_{-0.110}$ 
                   & $1.423^{+0.098}_{-0.111}$
                   & $1.409^{+0.100}_{-0.109}$
                   & $1.395^{+0.099}_{-0.108}$
                   & $1.387^{+0.103}_{-0.103}$
                   & $1.409^{+0.098}_{-0.110}$
                   & $1.397^{+0.097}_{-0.109}$
                   & $1.447^{+0.099}_{-0.111}$\\ 

$\beta_0$          & $1.406^{+0.394}_{-0.360}$
                   & $1.469^{+0.377}_{-0.376}$
                   & $1.336^{+0.413}_{-0.374}$
                   & $1.593^{+0.405}_{-0.368}$
                   & $1.437^{+0.411}_{-0.376}$
                   & $1.394^{+0.410}_{-0.368}$
                   & $1.449^{+0.409}_{-0.370}$
                   & $1.375^{+0.406}_{-0.371}$\\ 

$\beta_1$          & $5.219^{+0.978}_{-1.116}$
                   & $5.070^{+1.003}_{-1.126}$
                   & $5.395^{+1.009}_{-1.152}$
                   & $4.668^{+1.002}_{-1.141}$
                   & $5.085^{+1.017}_{-1.155}$
                   & $5.333^{+0.988}_{-1.153}$
                   & $5.035^{+0.989}_{-1.164}$
                   & $5.193^{+1.008}_{-1.134}$\\ 

$\Omega_{k0}$      & $-0.0009^{+0.0032}_{-0.0032}$
                   & $-0.0013^{+0.0038}_{-0.0044}$
                   & $-0.0108^{+0.0037}_{-0.0056}$
                   & $-0.0099^{+0.0039}_{-0.0056}$
                   & $-0.0118^{+0.0035}_{-0.0058}$
                   & $-0.0111^{+0.0035}_{-0.0052}$
                   & $-0.0118^{+0.0038}_{-0.0058}$
                   & $-0.0110^{+0.0037}_{-0.0052}$\\

$\Omega_{b0}$      & $0.0483^{+0.0015}_{-0.0015}$
                   & $0.0480^{+0.0022}_{-0.0024}$
                   & $0.0477^{+0.0022}_{-0.0025}$
                   & $0.0480^{+0.0022}_{-0.0025}$
                   & $0.0473^{+0.0022}_{-0.0024}$
                   & $0.0476^{+0.0022}_{-0.0025}$
                   & $0.0469^{+0.0022}_{-0.0024}$
                   & $0.0482^{+0.0022}_{-0.0023}$\\ 

$\Omega_{m0}$      & $0.300^{+0.010}_{-0.010}$
                   & $0.297^{+0.013}_{-0.015}$
                   & $0.295^{+0.013}_{-0.015}$
                   & $0.295^{+0.013}_{-0.015}$
                   & $0.294^{+0.013}_{-0.014}$
                   & $0.295^{+0.014}_{-0.014}$
                   & $0.290^{+0.013}_{-0.013}$
                   & $0.299^{+0.013}_{-0.014}$\\ 

$h$                & $0.684^{+0.010}_{-0.010}$
                   & $0.687^{+0.015}_{-0.015}$
                   & $0.689^{+0.016}_{-0.016}$
                   & $0.687^{+0.015}_{-0.016}$
                   & $0.691^{+0.016}_{-0.016}$
                   & $0.689^{+0.016}_{-0.016}$
                   & $0.695^{+0.016}_{-0.016}$
                   & $0.685^{+0.016}_{-0.016}$\\ 
$w_0$              & $...$
                   & $-1.014^{+0.080}_{-0.080}$
                   & $-0.705^{+0.207}_{-0.212}$
                   & $-0.648^{+0.252}_{-0.252}$
                   & $-0.815^{+0.177}_{-0.176}$
                   & $-0.764^{+0.185}_{-0.181}$
                   & $-0.690^{+0.203}_{-0.204}$
                   & $-0.710^{+0.203}_{-0.205}$\\ 

$w_a$              & $...$
                   & $...$
                   & $-2.286^{+1.675}_{-1.469}$
                   & $-3.419^{+2.290}_{-2.286}$
                   & $-1.200^{+0.905}_{-0.880}$
                   & $8.405^{+5.688}_{-6.328}$
                   & $-2.335^{+1.291}_{-1.278}$
                   & $-1.446^{+0.359}_{-0.292}$\\ 

$\chi^2_{min}$     & $386.5$
                   & $386.5$
                   & $383.7$
                   & $384.1$
                   & $383.8$
                   & $383.8$
                   & $383.7$
                   & $384.0$\\ 
\hline
\end{tabular}
\end{table*}

For the theoretical side,
let us discuss the impacts of different $w(z)$ first.
Table~\ref{tab:res_betaz_cmb1_bao1} summarizes the fitting results of the six DE parametrization models.
To make a comparison, we also list the fitting results of the $\Lambda$CDM model and the $w$CDM model.
Notice that a non-flat Universe is adopted in the background.
From this table we see that, all the six DE parametrization models give a similar value of $\chi^2_{min}$,
which is smaller that the $\chi^2_{min}$ values of the $\Lambda$CDM and the $w$CDM model by $\sim 2.5$.
Besides, for these six models,
the differences of various cosmological parameters (such as $\Omega_{m0}$, $\Omega_{b0}$, and $h$) are very small.
This means that current observations can not distinguish these six dynamical DE models,
and thus can not put tight constraints on the evolution of EoS $w(z)$.
We also find that, for the $\Lambda$CDM and $w$CDM models,
the curvature component $\Omega_{k0}=0$ (i.e., consistent with a flat Universe) at $1\sigma$ CL,
which is consistent with the results of many previous studies
~\citep{Padmanabhan2012, Anderson2012, Anderson2014a, Anderson2014b, Sanchez2014, Samushia2014, Kazin2014, Planck201513}.
However, after taking into account the evolution of EoS $w$,
the curvature component $\Omega_{k0}$ will deviate from 0 at $2\sigma$ CL.
This implies that the spatial curvature has a degeneracy with the evolution of EoS $w(z)$,
and then shows the importance of considering the effects of spatial curvature on CA.
In addition, we find that, all the six dynamical DE models give a current EoS value $w_0$ deviating from $-1$ at 1$\sigma$ CL.
Moreover, the MZ model give a $w_a$ larger than $0$ at 1$\sigma$ CL,
while other five parametrization models give a $w_a$ smaller than $0$ at 1$\sigma$ CL;
as will be seen below, these constraints on $w_a$ will lead to a slowing down CA.
\footnote{For the MZ model, $\frac{ln(2+z)}{1+z}$ is a decreasing function of redshift $z$,
so a $w_a$ larger than $0$ will lead to a slowing down CA.}

\begin{figure*}
  \centering
  \resizebox{0.74\columnwidth}{!}{\includegraphics{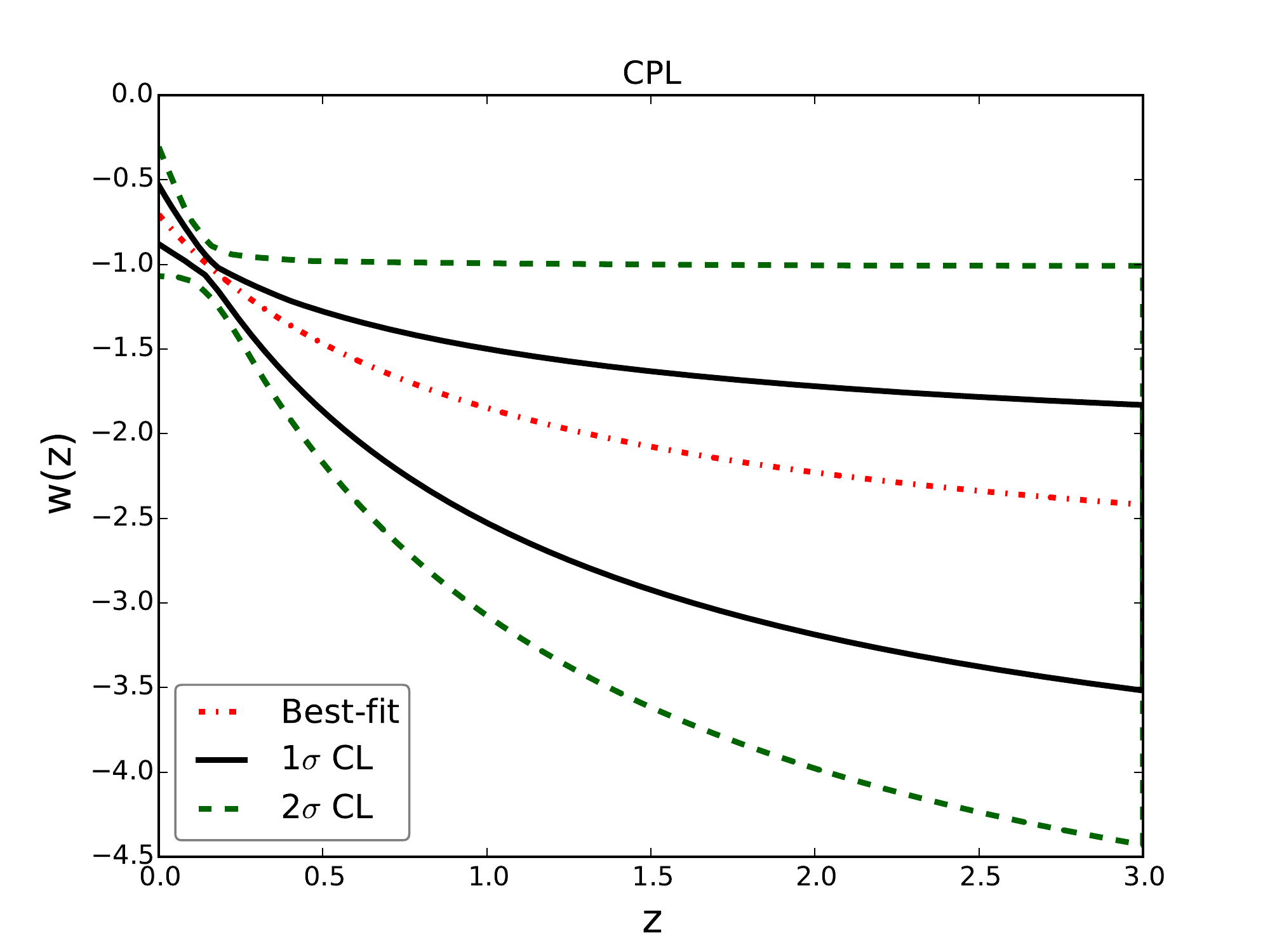}}
  \hspace{0.1\columnwidth}
  \resizebox{0.74\columnwidth}{!}{\includegraphics{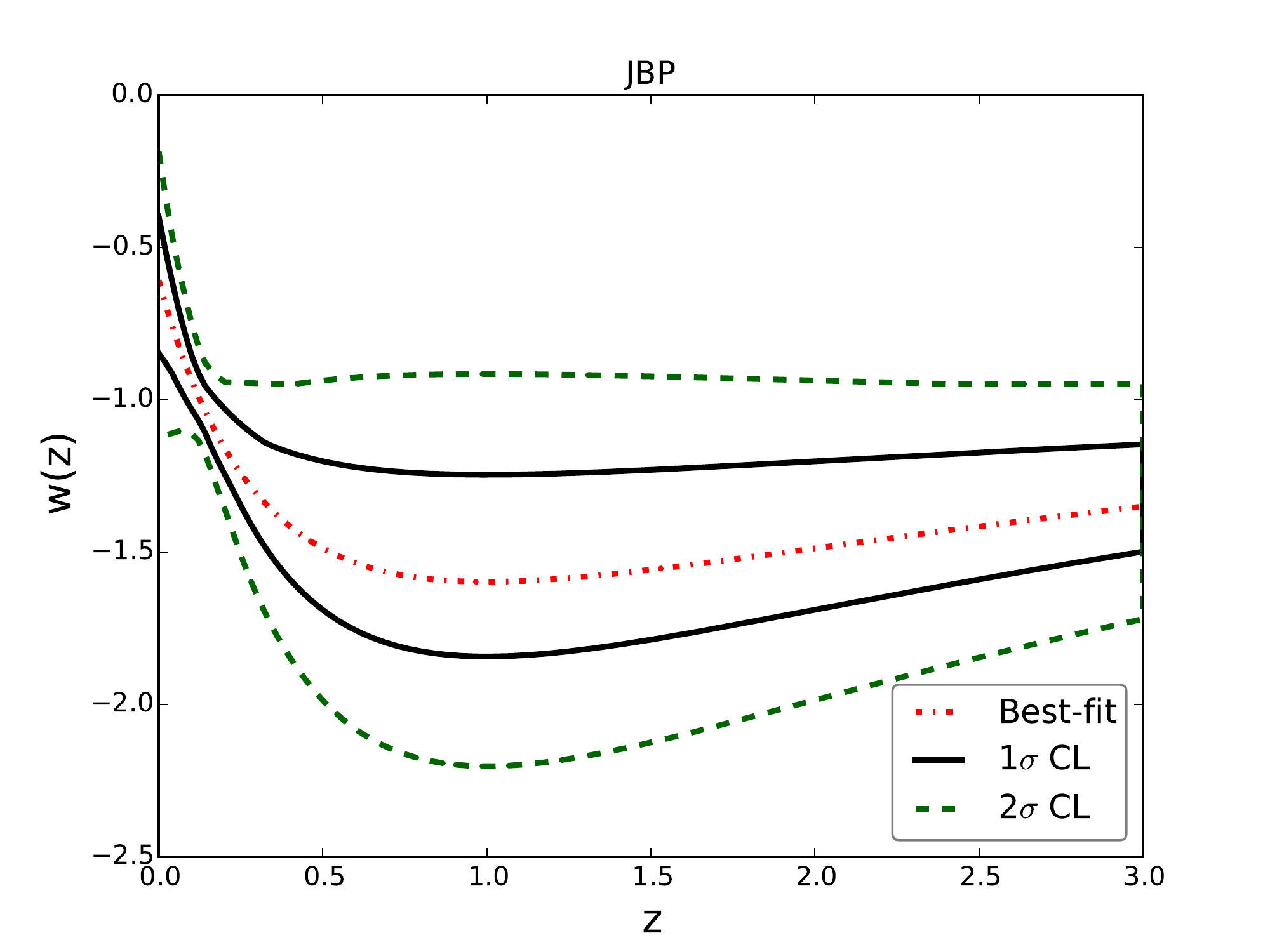}}
  \hspace{0.1\columnwidth}
  \resizebox{0.74\columnwidth}{!}{\includegraphics{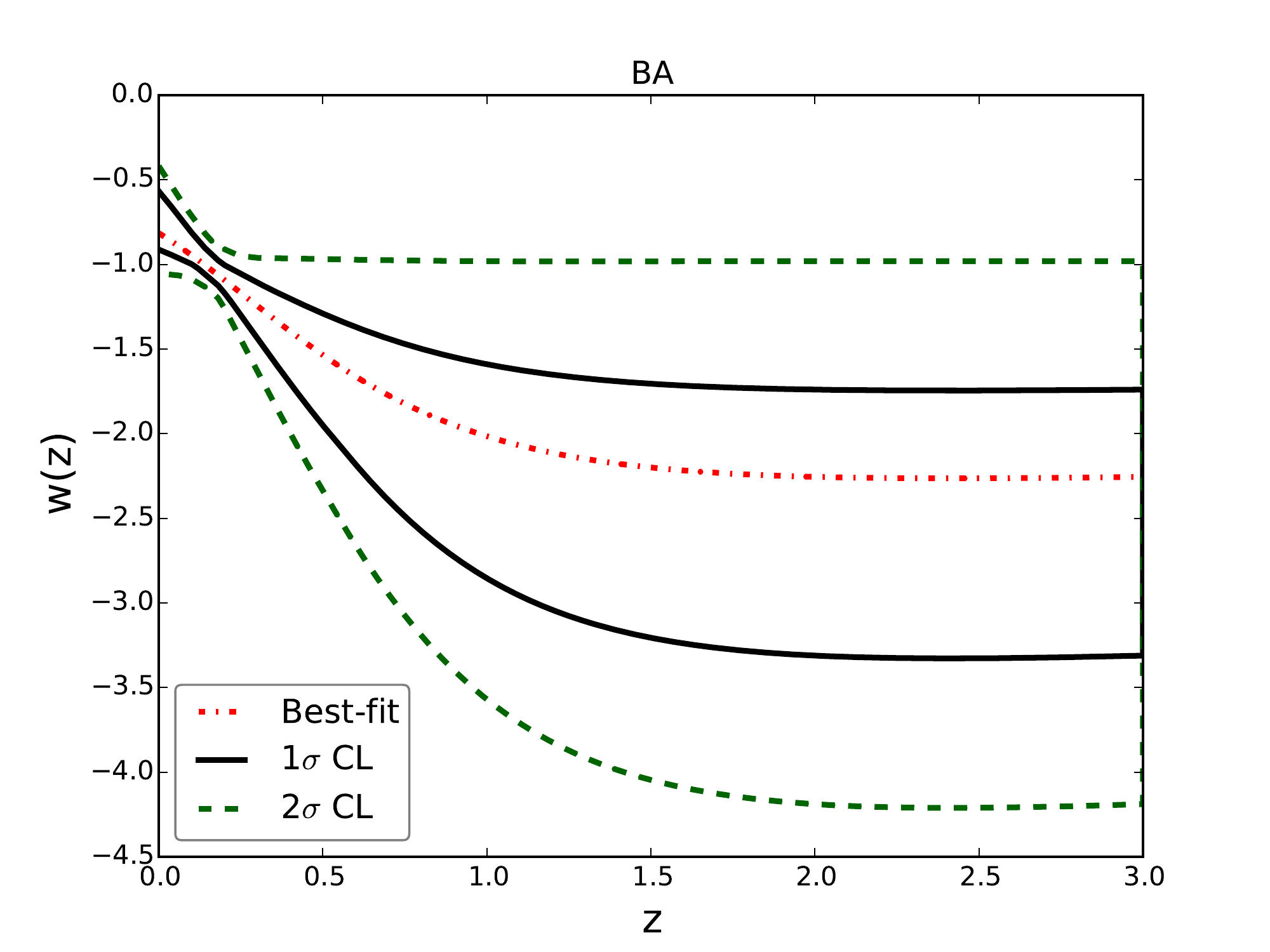}}
  \hspace{0.1\columnwidth}
  \resizebox{0.74\columnwidth}{!}{\includegraphics{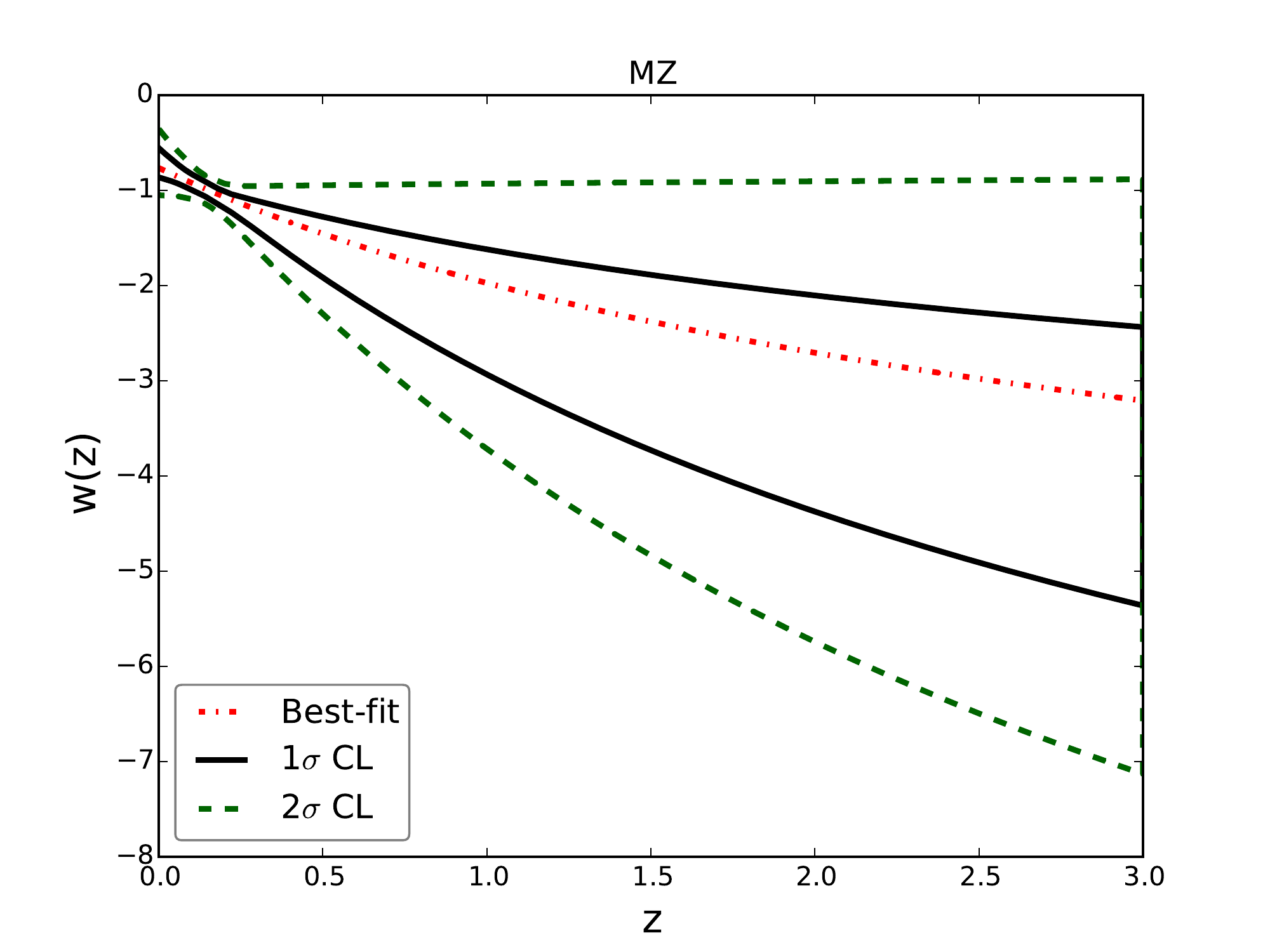}}
  \hspace{0.1\columnwidth}
  \resizebox{0.74\columnwidth}{!}{\includegraphics{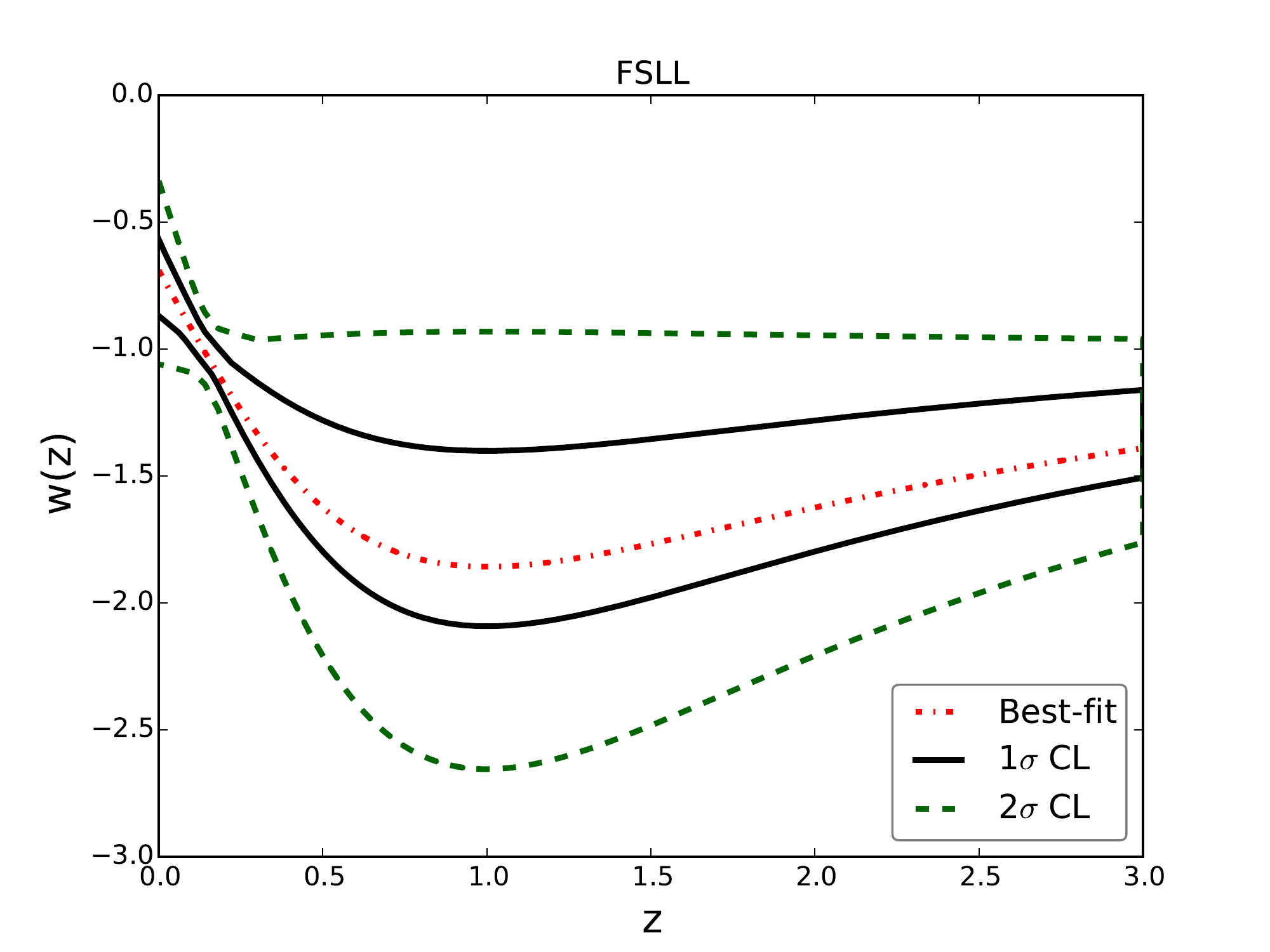}}
  \hspace{0.1\columnwidth}
  \resizebox{0.74\columnwidth}{!}{\includegraphics{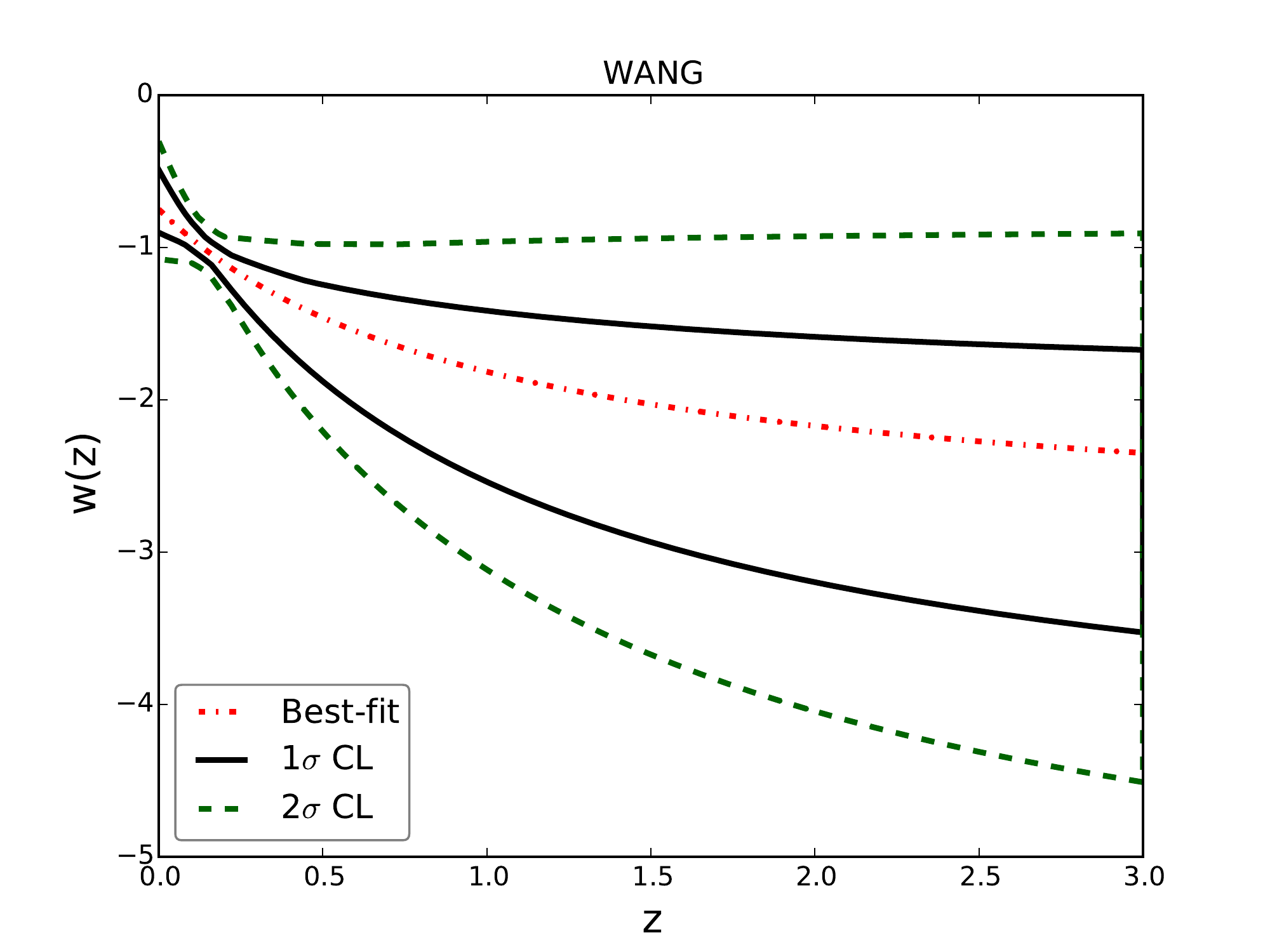}}
  \caption{(color online). The best-fit results (red dash-dotted line), the 1$\sigma$ confidence regions (black solid lines) and the 2$\sigma$ confidence regions (green dashed lines) of EoS $w(z)$ for the six DE parametrization models.
}
\label{fig:o_wz_betaz_cmb1_bao1}
\end{figure*}

Now we discuss this issue with more details.
In Fig.~\ref{fig:o_wz_betaz_cmb1_bao1}, we plot the best-fit results, the 1$\sigma$ and the 2$\sigma$ confidence regions of $w(z)$
for the six DE parametrization models.
We find that, for all the parametrization models, EoS $w$ deviates from a cosmological constant at 1$\sigma$ CL,
which shows the importance of considering the evolution of EoS.
Moreover, although different DE model gives a different evolutionary behavior of $w(z)$ at high redshift,
all the six DE models give an increasing $w(z)$ when $z \rightarrow 0$ at almost 2$\sigma$ CL.
This indicates that all the dynamical DE models favor a slowing down CA.

\begin{figure*}
  \centering
  \resizebox{0.74\columnwidth}{!}{\includegraphics{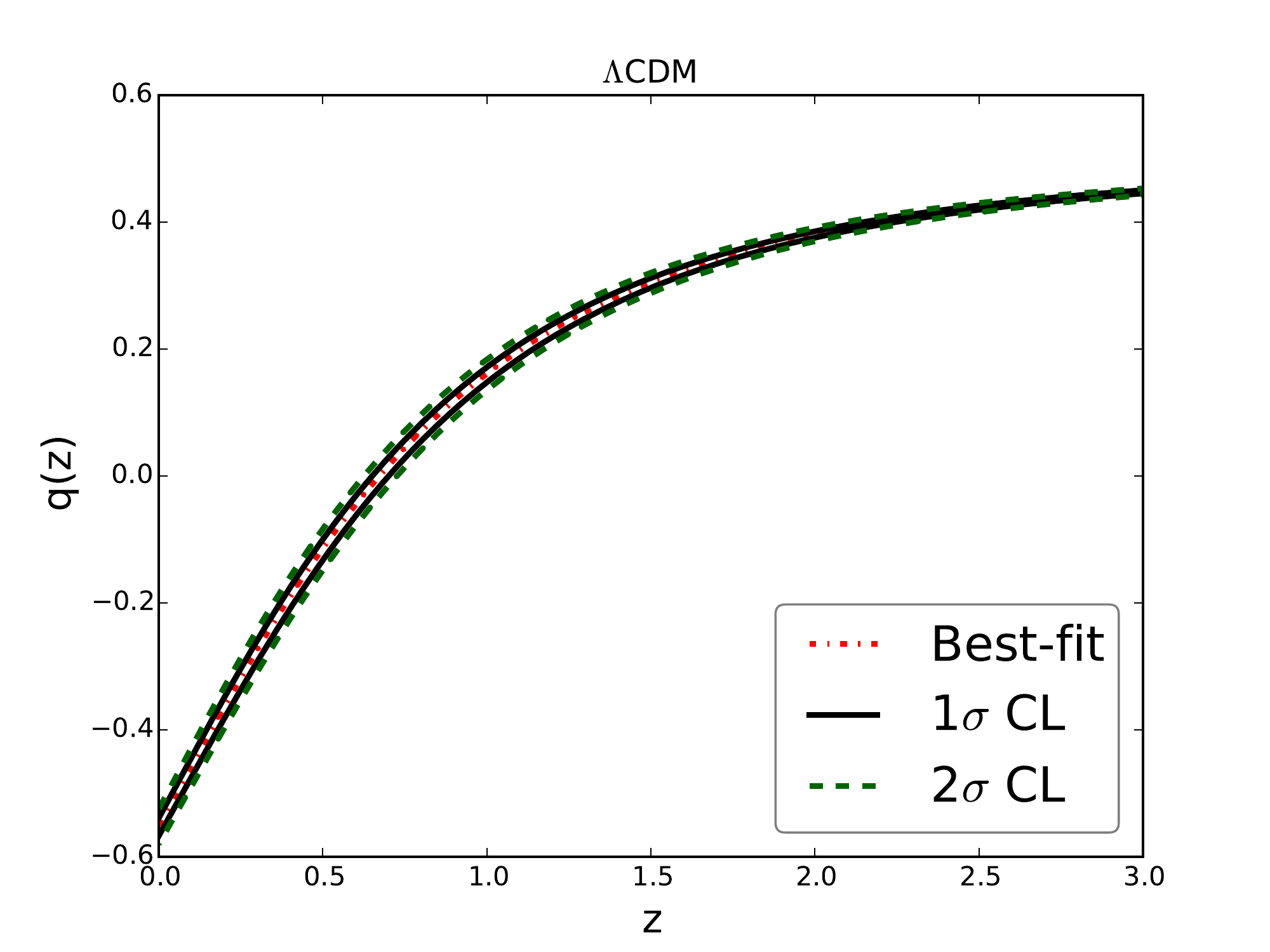}}
  \hspace{0.1\columnwidth}
  \resizebox{0.74\columnwidth}{!}{\includegraphics{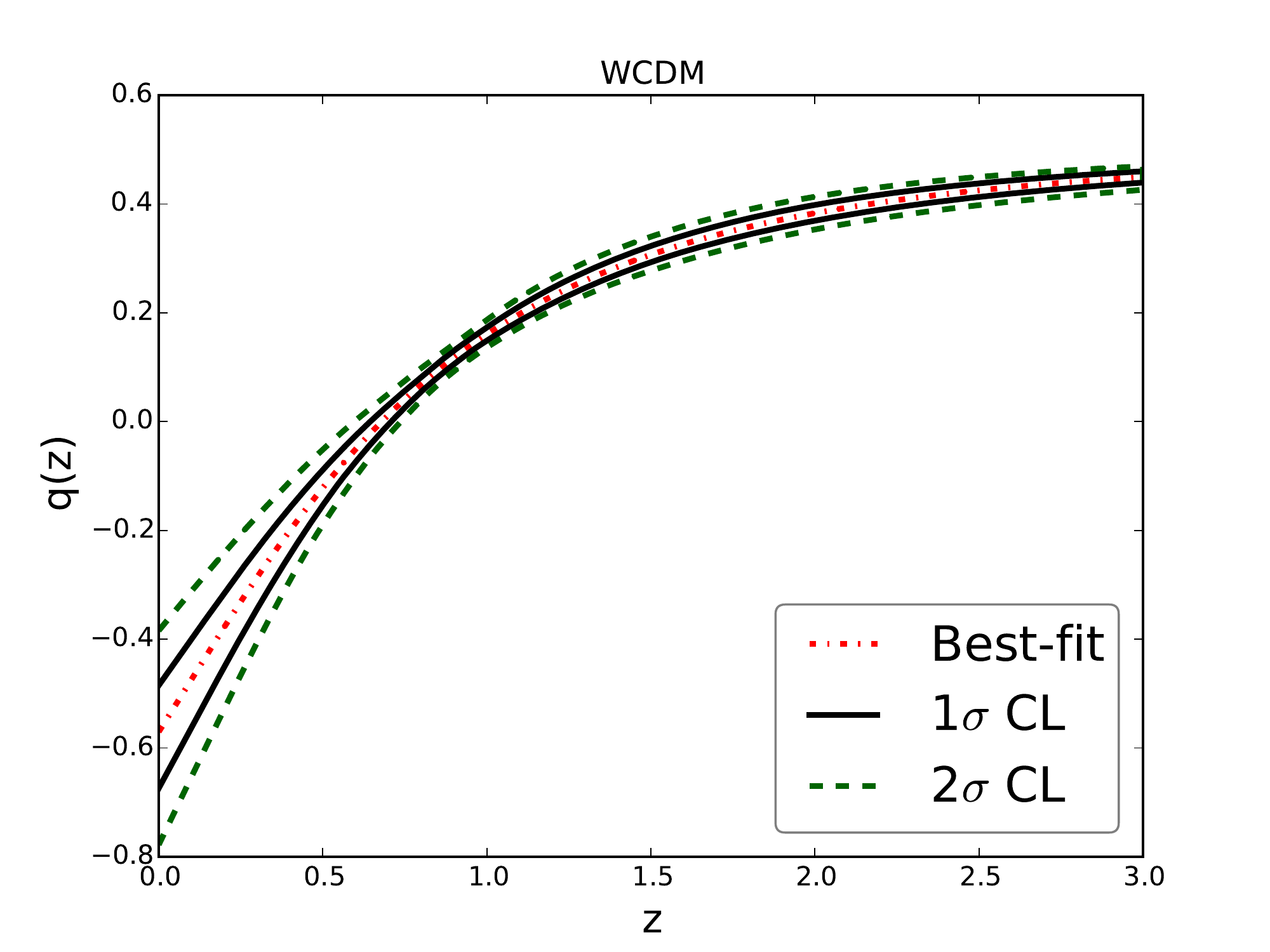}}
  \hspace{0.1\columnwidth}
  \resizebox{0.74\columnwidth}{!}{\includegraphics{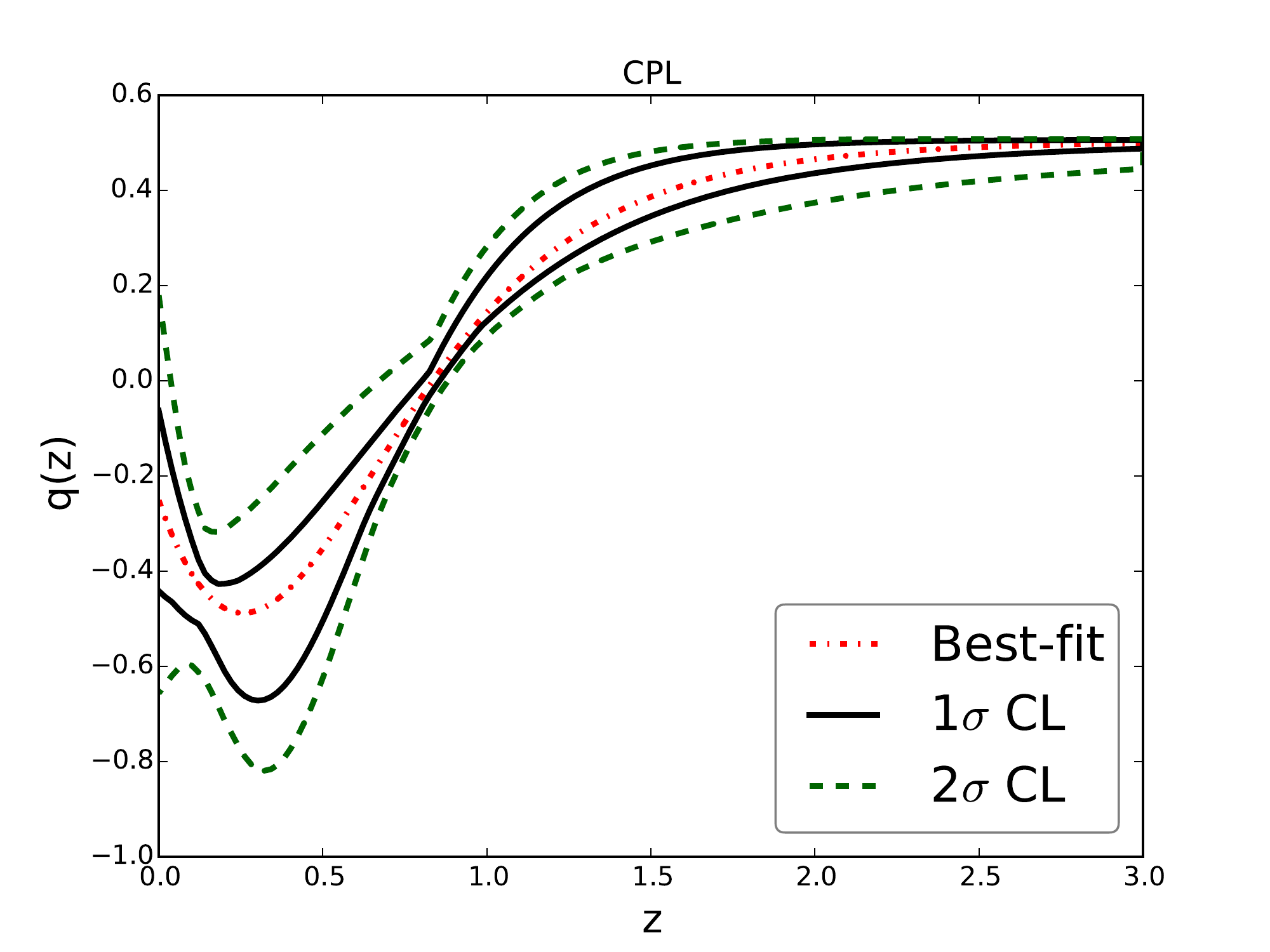}}
  \hspace{0.1\columnwidth}
  \resizebox{0.74\columnwidth}{!}{\includegraphics{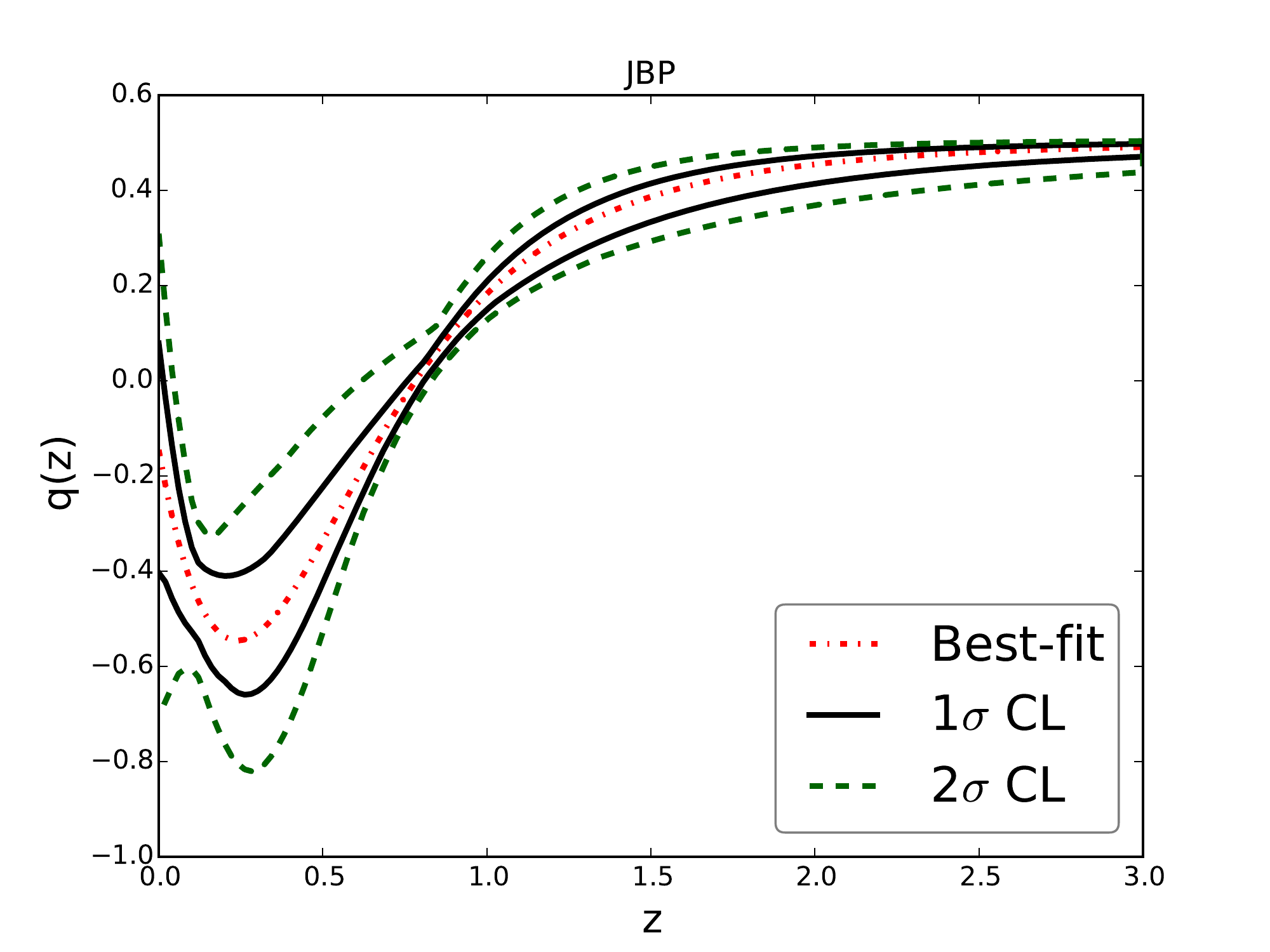}}
  \hspace{0.1\columnwidth}
  \resizebox{0.74\columnwidth}{!}{\includegraphics{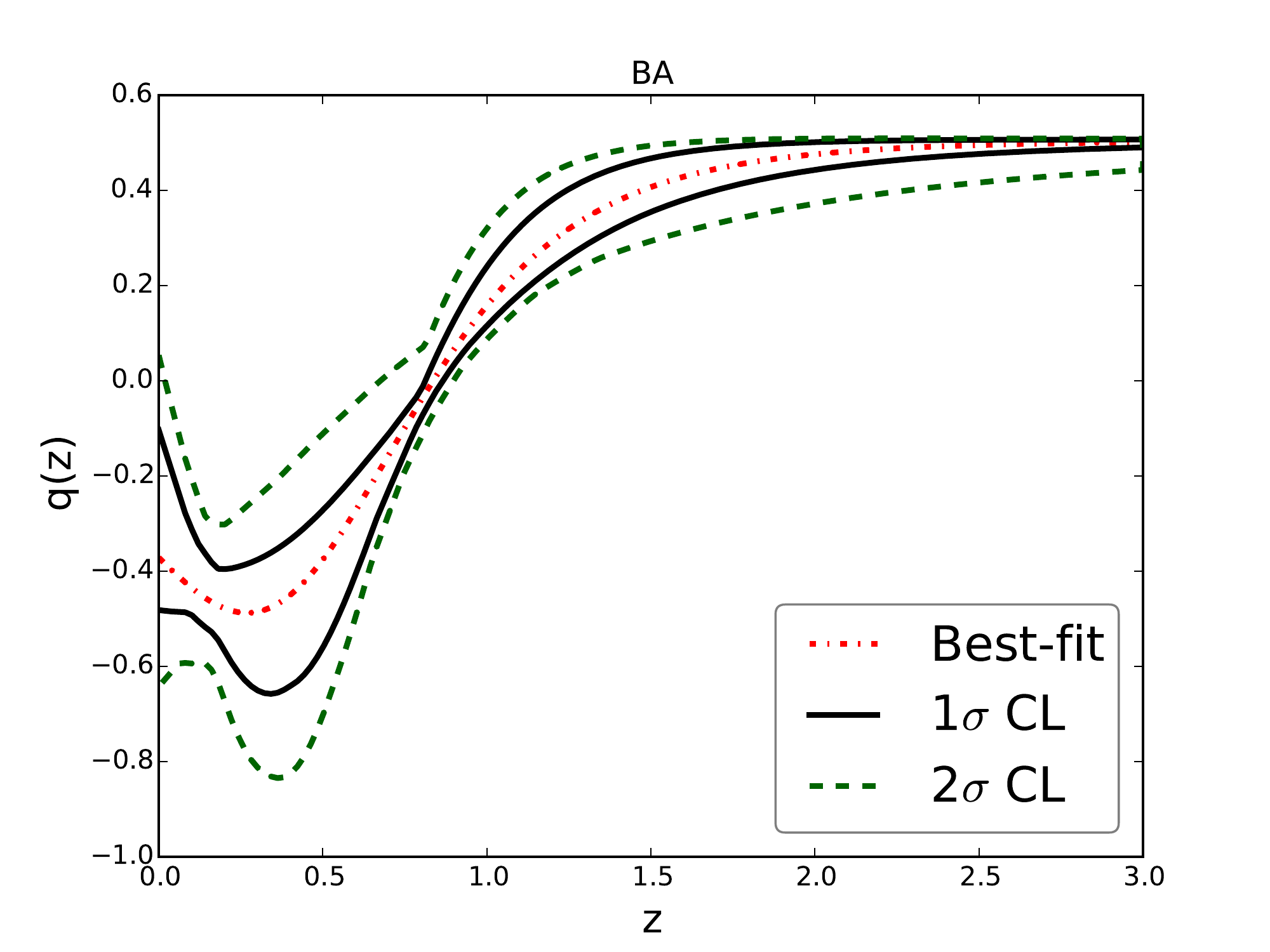}}
  \hspace{0.1\columnwidth}
  \resizebox{0.74\columnwidth}{!}{\includegraphics{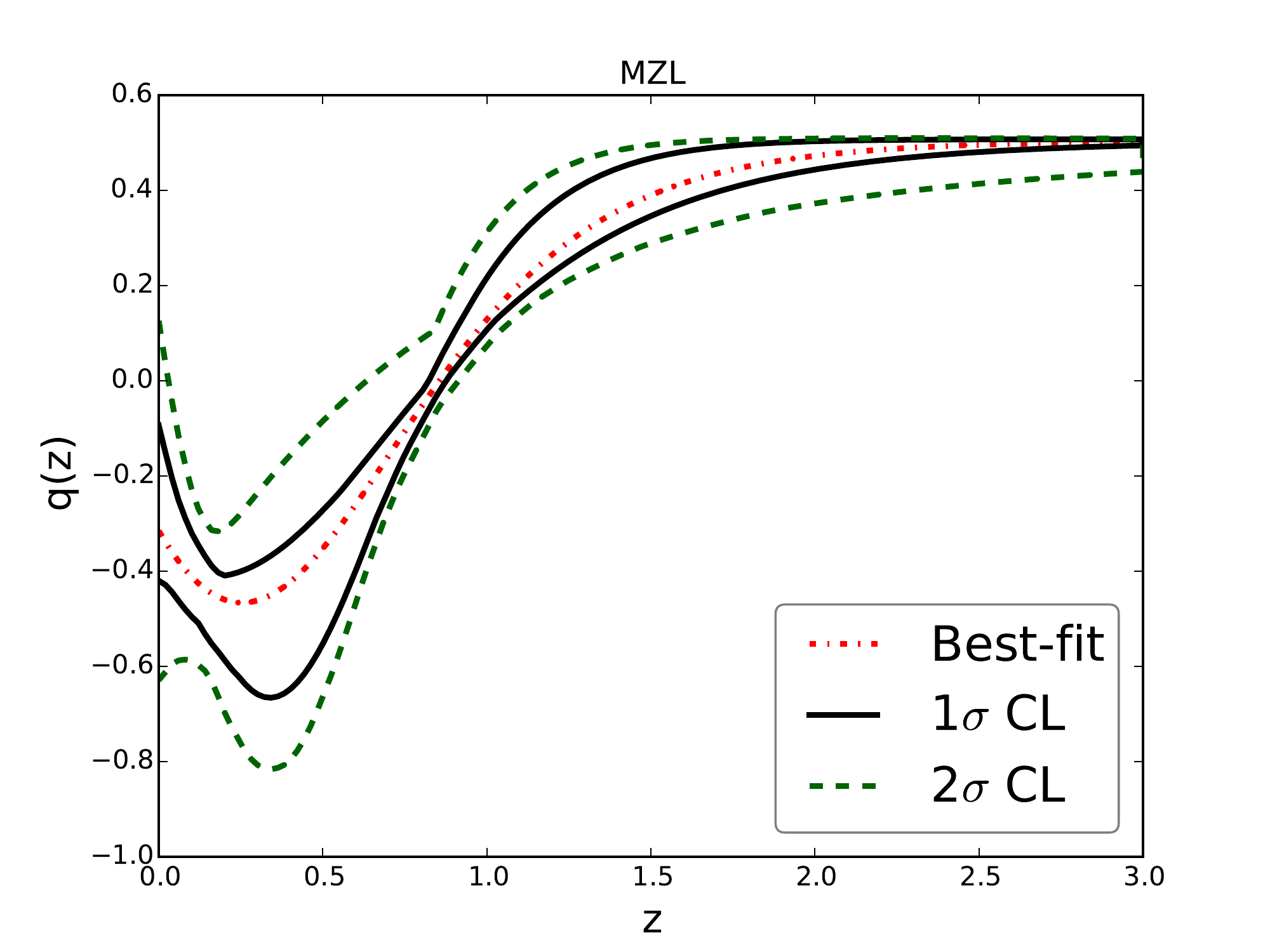}}
  \hspace{0.1\columnwidth}
  \resizebox{0.74\columnwidth}{!}{\includegraphics{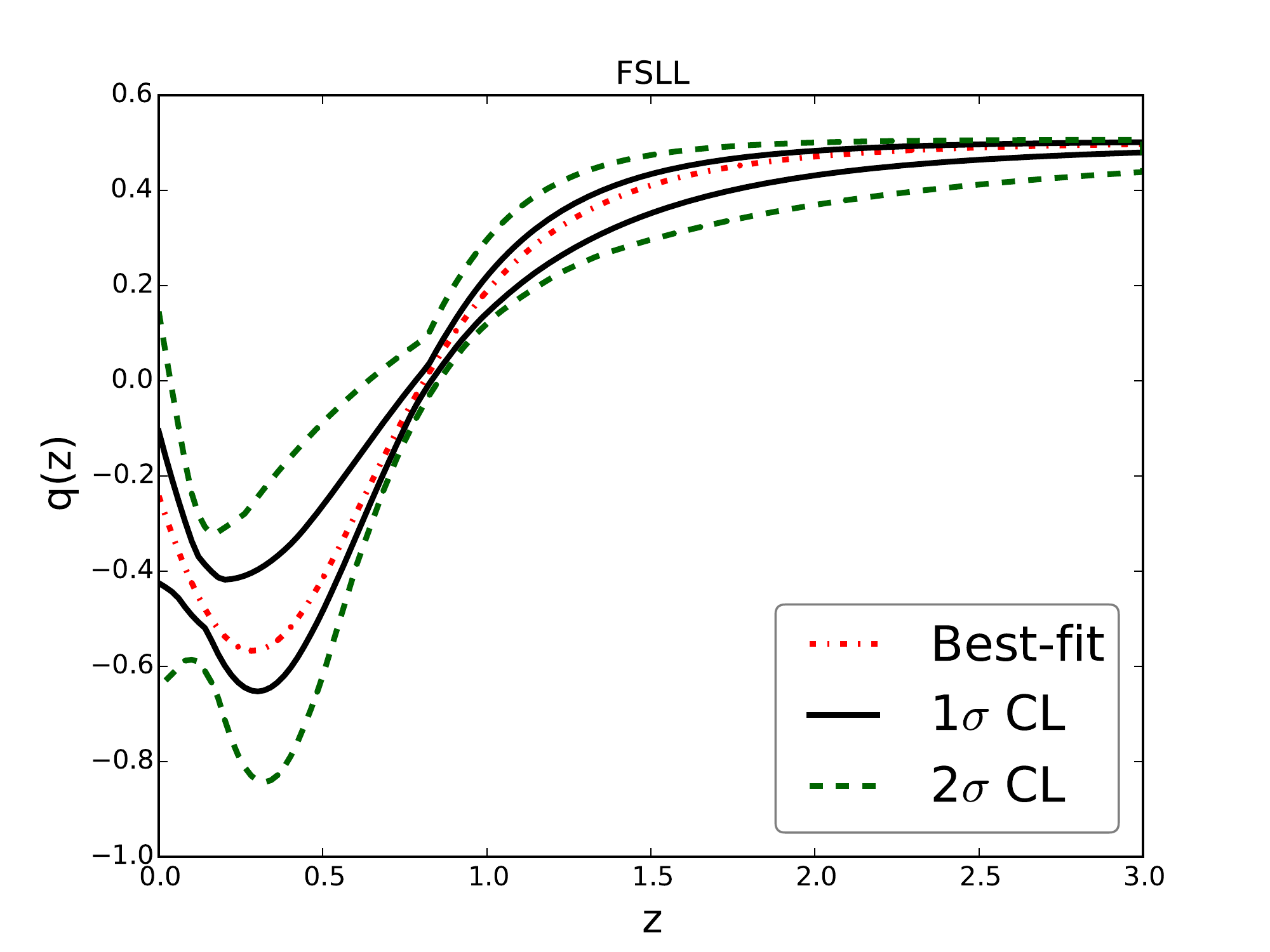}}
  \hspace{0.1\columnwidth}
  \resizebox{0.74\columnwidth}{!}{\includegraphics{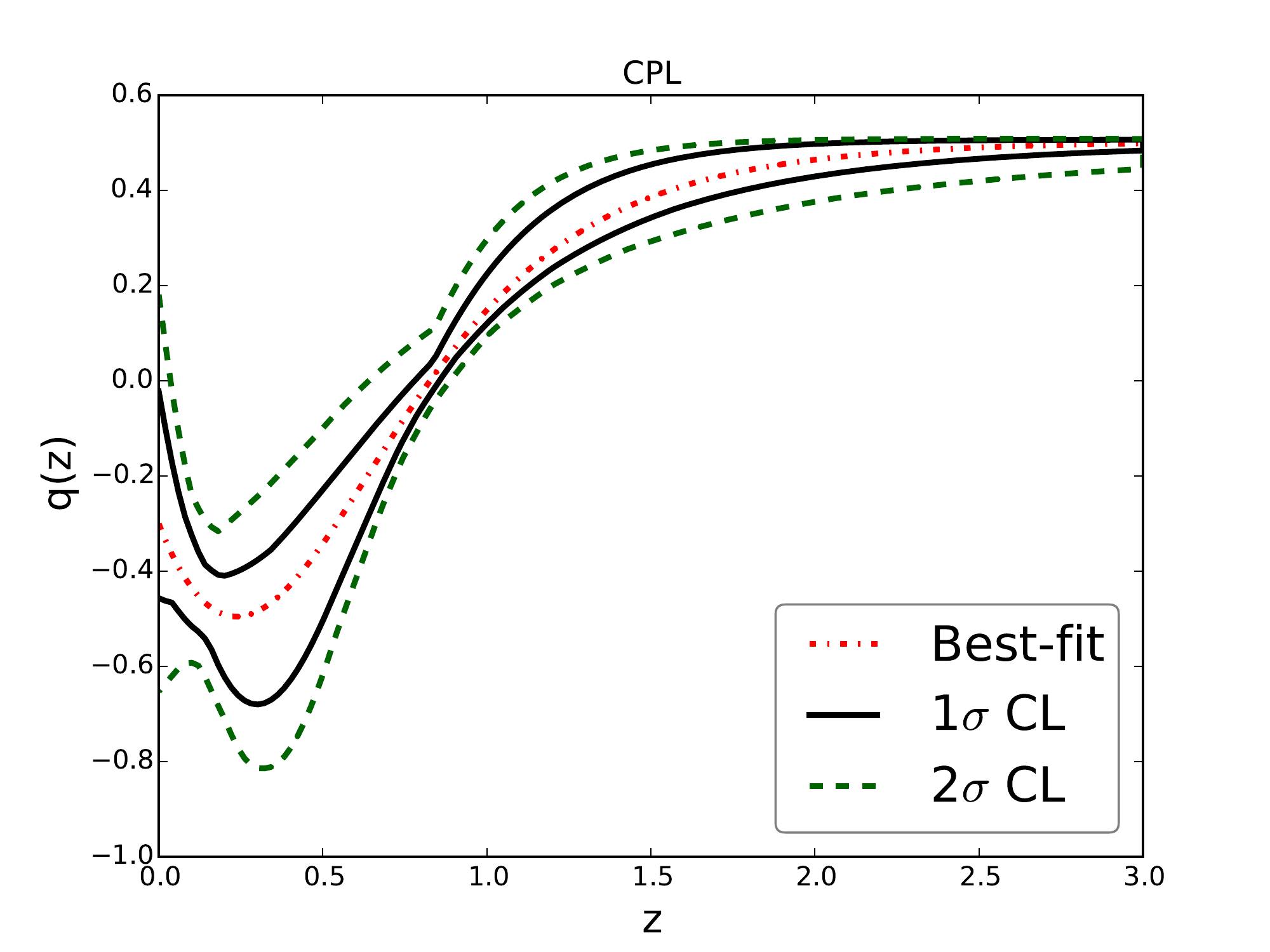}}
  \caption{(color online). The best-fit results (red dash-dotted line), the 1$\sigma$ confidence regions (black solid lines) and the 2$\sigma$ confidence regions (green dashed lines) of deceleration parameter $q(z)$ for the $\Lambda$CDM model, the $w$CDM model and the six DE parametrization models.
}
\label{fig:o_qz_betaz_cmb1_bao1}
\end{figure*}

Moreover, we plot the best-fit results, the 1$\sigma$ and the 2$\sigma$ confidence regions of $q(z)$
for the six DE parametrization models in Fig.~\ref{fig:o_qz_betaz_cmb1_bao1}.
To make a comparison, we also give the corresponding results of the $\Lambda$CDM and the $w$CDM model.
It is clear that the $\Lambda$CDM and the $w$CDM model give a decreasing $q(z)$ when $z \rightarrow 0$,
which means that these two models prefer an eternal CA.
In contrast, for the six DE parametrization models, $q(z)$ achieves its bottom at redshift $z\sim0.26$,
and then becomes a increasing function of $z$ at more than 1$\sigma$ CL.
This implies that all the dynamical DE models favor a slowing down CA.

\begin{table*}
\centering
\caption{The $\rm \chi^2_{min}$s, $\rm \Delta AIC$s and $\rm \Delta BIC$s of the $\rm \Lambda CDM$ and the six parametrization models.
}
\label{tab:aicbic}
\centering
\begin{tabular}{ccccccccccc}
\hline\hline 
Parameter  & $\Lambda$CDM & CPL & JBP & BA & MZ & FSLL & WANG \\ \hline
$\rm \chi^2_{min}$              & $386.5$
                                & $383.7$
                                & $384.1$
                                & $383.3$
                                & $383.8$
                                & $383.7$
                                & $384.0$\\ 

$\rm \Delta AIC$                & $0$
                                & $1.2$
                                & $1.6$
                                & $1.3$
                                & $1.3$
                                & $1.2$
                                & $1.5$\\ 

$\rm \Delta BIC$                & $0$
                                & $9.5$
                                & $9.9$
                                & $9.6$
                                & $9.6$
                                & $9.5$
                                & $9.8$\\ 
\hline
\end{tabular}
\end{table*}

To give a quantitative description on the transition from an eternal CA phase to a slowing down CA phase.
Here we use the transition redshift $z_p$ where the deceleration parameter $q(z)$ achieve its bottom.
In Table~\ref{tab:res_z_peak}, we list the best-fit values of $z_p$ given by the six DE parametrization models.
It can be seen that all the six DE model undergo a phase transition at the redshift region $[0.23,0.29]$;
besides, the FSLL model gives a largest $z_p$, while the WANG model gives a smallest $z_p$.

\begin{table*}
\centering
\caption{Best-fit results of transition redshift $z_p$ given by the six parametrization models.
}
\label{tab:res_z_peak}
\centering
\begin{tabular}{ccccccccccc}
\hline\hline 
Parameter  & CPL & JBP & BA & MZ & FSLL & WANG \\ \hline
$z_p$              & $0.258$
                   & $0.241$
                   & $0.269$
                   & $0.254$
                   & $0.286$
                   & $0.238$\\ 
\hline
\end{tabular}
\end{table*}

Moreover, to assess the statistical significance of the slowing down trend of CA,
we adopt the Akaike information criteria (AIC)~\citep{AIC} and Bayesian information criteria (BIC)~\citep{BIC}
that are defined as
\begin{equation}
  {\rm AIC} = \chi^2_{min}+2k, \ \ {\rm BIC} = \chi^2_{min}+k \ln N,
\end{equation}
where $k$ is the number of free parameters, and $N$ is the number of total data points.
In Table~\ref{tab:aicbic}, we list the results of $\chi^2_{min}$, $\Delta {\rm AIC}_{\rm model}$ and $\Delta {\rm BIC}_{\rm model}$ for
the $\Lambda$CDM and the six DE parametrization models, where
\ba
\Delta {\rm AIC}_{\rm model}\equiv {\rm AIC}_{\rm model}-{\rm AIC}_{\rm \Lambda CDM},  \nonumber \\
\Delta {\rm BIC}_{\rm model}\equiv {\rm BIC}_{\rm model}-{\rm BIC}_{\rm \Lambda CDM}.
\label{eq:ic}
\ea
From this table we see that, compared with the $\Lambda$CDM model, all the six DE parametrization models yield larger AICs and BICs.
This means that the dynamical evolution of EoS is not favored by the current observational data.

In conclusion, once the dynamical evolution of DE EoS $w$ is taken into account,
the combined $\rm SNLS3(linear\ \beta)+BAO(1D)+Planck2015$ data prefer a slowing down CA at more than 1$\sigma$ CL.
Since all the six DE models give same evolutionary trends of $w(z)$ and $q(z)$ when $z \rightarrow 0$,
our conclusion is insensitive to the specific functional form of $w(z)$.
This conclusion is consistent with the results of Ref.~\citep{Magana2014}.
In addition, due to low statistical significance, the slowing down of CA is still a theoretical possibility that cannot confirmed by the current observations.

\subsubsection{Impacts of Spatial Curvature}
\label{subsubsec:spatial curvature}

\begin{table*}
\centering
\caption{Fitting results of the CPL, the JBP and the FSLL models, where both the best-fit values and the $1\sigma$ errors of various parameters are listed.
The legends `` WITH $\Omega_{k0}$'' and `` WITHOUT $\Omega_{k0}$'' represent the cases with and without considering spatial curvature, respectively.}
\label{tab:res_curv_compare}
\centering
\begin{tabular}{ccccccccccccc}
\hline\hline &\multicolumn{2}{c}{CPL}&&\multicolumn{2}{c}{JBP}&&\multicolumn{2}{c}{FSLL} \\
           \cline{2-3}\cline{5-6}\cline{8-9}
           Parameter  &  WITH $\Omega_{k0}$ &  WITHOUT $\Omega_{k0}$ & &  WITH $\Omega_{k0}$ &  WITHOUT $\Omega_{k0}$ & &  WITH $\Omega_{k0}$ &  WITHOUT $\Omega_{k0}$    \\ \hline
$\alpha$           & $1.409^{+0.100}_{-0.109}$
                   & $1.421^{+0.099}_{-0.110}$&
                   & $1.395^{+0.099}_{-0.108}$
                   & $1.384^{+0.098}_{-0.110}$&
                   & $1.397^{+0.097}_{-0.109}$
                   & $1.413^{+0.099}_{-0.111}$\\ 

$\beta_0$          & $1.336^{+0.413}_{-0.374}$
                   & $1.466^{+0.402}_{-0.371}$&
                   & $1.593^{+0.405}_{-0.368}$
                   & $1.437^{+0.403}_{-0.374}$&
                   & $1.449^{+0.409}_{-0.370}$
                   & $1.466^{+0.411}_{-0.374}$\\ 

$\beta_1$          & $5.395^{+1.009}_{-1.152}$
                   & $5.129^{+1.001}_{-1.145}$&
                   & $4.668^{+1.002}_{-1.141}$
                   & $5.226^{+1.011}_{-1.139}$&
                   & $5.035^{+0.989}_{-1.164}$
                   & $4.975^{+1.007}_{-1.161}$\\ 

$\Omega_{k0}$      & $-0.0108^{+0.0037}_{-0.0056}$
                   & $...$&
                   & $-0.0099^{+0.0039}_{-0.0056}$
                   & $...$&
                   & $-0.0118^{+0.0038}_{-0.0058}$
                   & $...$\\

$\Omega_{b0}$      & $0.0477^{+0.0022}_{-0.0025}$
                   & $0.0479^{+0.0023}_{-0.0024}$&
                   & $0.0480^{+0.0022}_{-0.0025}$
                   & $0.0477^{+0.0023}_{-0.0023}$&
                   & $0.0469^{+0.0022}_{-0.0024}$
                   & $0.0474^{+0.0023}_{-0.0025}$\\ 

$\Omega_{m0}$      & $0.295^{+0.013}_{-0.015}$
                   & $0.300^{+0.014}_{-0.014}$&
                   & $0.295^{+0.013}_{-0.015}$
                   & $0.298^{+0.013}_{-0.014}$&
                   & $0.290^{+0.013}_{-0.013}$
                   & $0.295^{+0.013}_{-0.015}$\\ 

$h$                & $0.689^{+0.016}_{-0.016}$
                   & $0.686^{+0.016}_{-0.016}$&
                   & $0.687^{+0.015}_{-0.016}$
                   & $0.688^{+0.016}_{-0.016}$&
                   & $0.695^{+0.016}_{-0.016}$
                   & $0.691^{+0.016}_{-0.016}$\\ 

$w_0$              & $-0.705^{+0.207}_{-0.212}$
                   & $-0.982^{+0.134}_{-0.134}$&
                   & $-0.648^{+0.252}_{-0.252}$
                   & $-0.960^{+0.181}_{-0.179}$&
                   & $-0.690^{+0.203}_{-0.204}$
                   & $-0.968^{+0.145}_{-0.144}$\\ 

$w_a$              & $-2.286^{+1.675}_{-1.469}$
                   & $-0.082^{+0.655}_{-0.440}$&
                   & $-3.419^{+2.290}_{-2.286}$
                   & $-0.317^{+1.341}_{-1.149}$&
                   & $-2.335^{+1.291}_{-1.278}$
                   & $-0.165^{+0.641}_{-0.527}$\\ 

$\chi^2_{min}$     & $383.7$
                   & $386.6$&
                   & $384.1$
                   & $386.6$&
                   & $383.7$
                   & $386.6$\\ 
\hline
\end{tabular}
\end{table*}

Then, let us turn to the impacts of spatial curvature.
Table~\ref{tab:res_curv_compare} shows the fitting results of the CPL, JBP and FSLL models,
where both the cases with and without spatial curvature are taken into account.
We find that, in a flat Universe,
these three parametrization models give a $w_0$ consistent with $-1$ and a $w_a$ consistent with $0$ at 1$\sigma$ CL.
In contrast, in a non-flat Universe, $w_0$ deviates from $-1$ at 1$\sigma$ CL, while $w_a$ deviates from $0$ at 1$\sigma$ CL.
This means that there is a significant degeneracy between the spatial curvature and the EoS $w(z)$,
which is consistent with the results of~\citep{Clarkson2007, Anderson2014a, Anderson2014b, Sanchez2014, Kazin2014}.
In addition, it is clear that this conclusion is insensitive to the specific functional form of $w(z)$.
These results show the importance of considering spatial curvature in the study of CA.

\begin{figure*}
  \centering
  \resizebox{0.70\columnwidth}{!}{\includegraphics{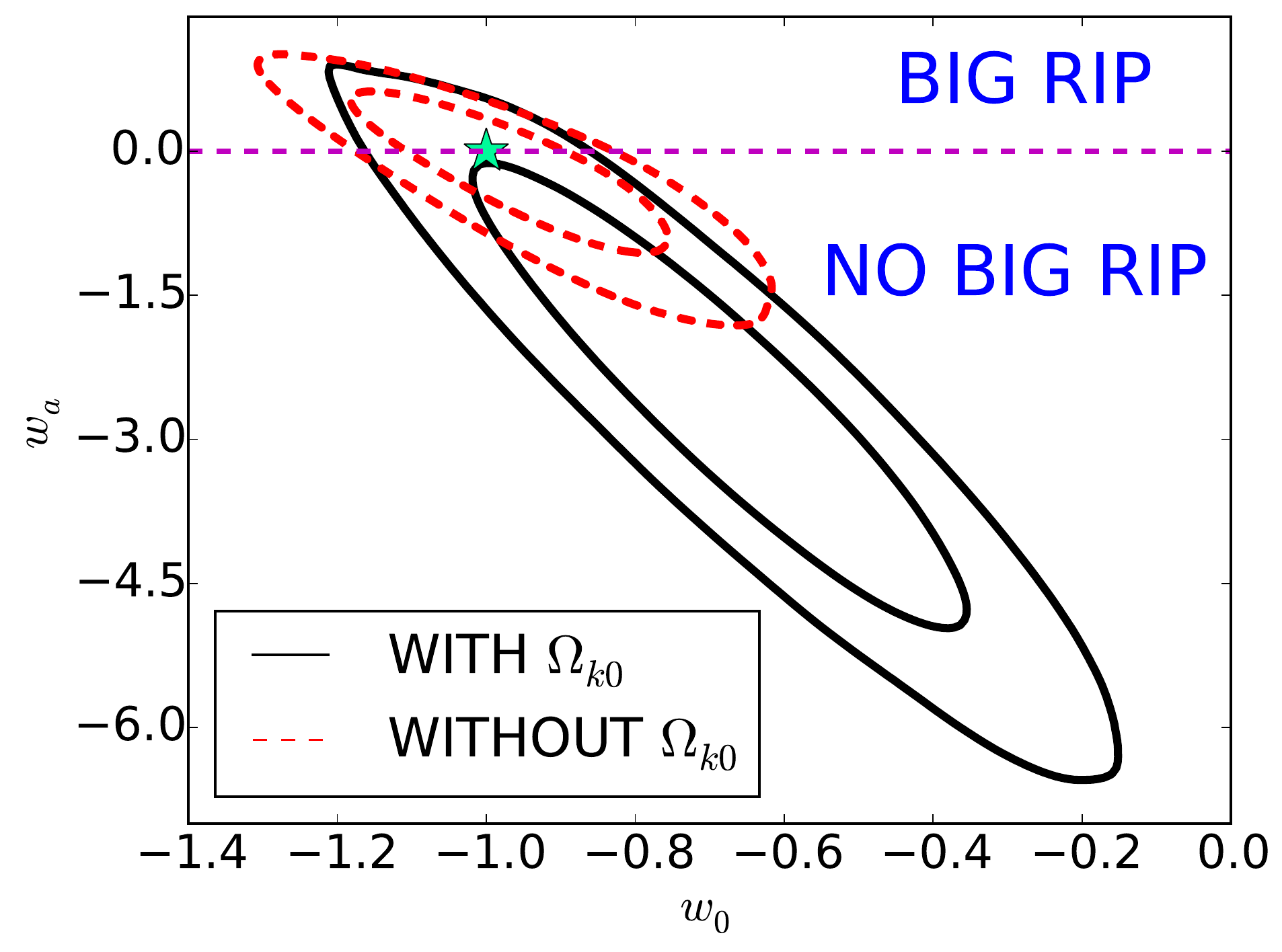}}
  \hspace{0.1\columnwidth}
  \resizebox{0.74\columnwidth}{!}{\includegraphics{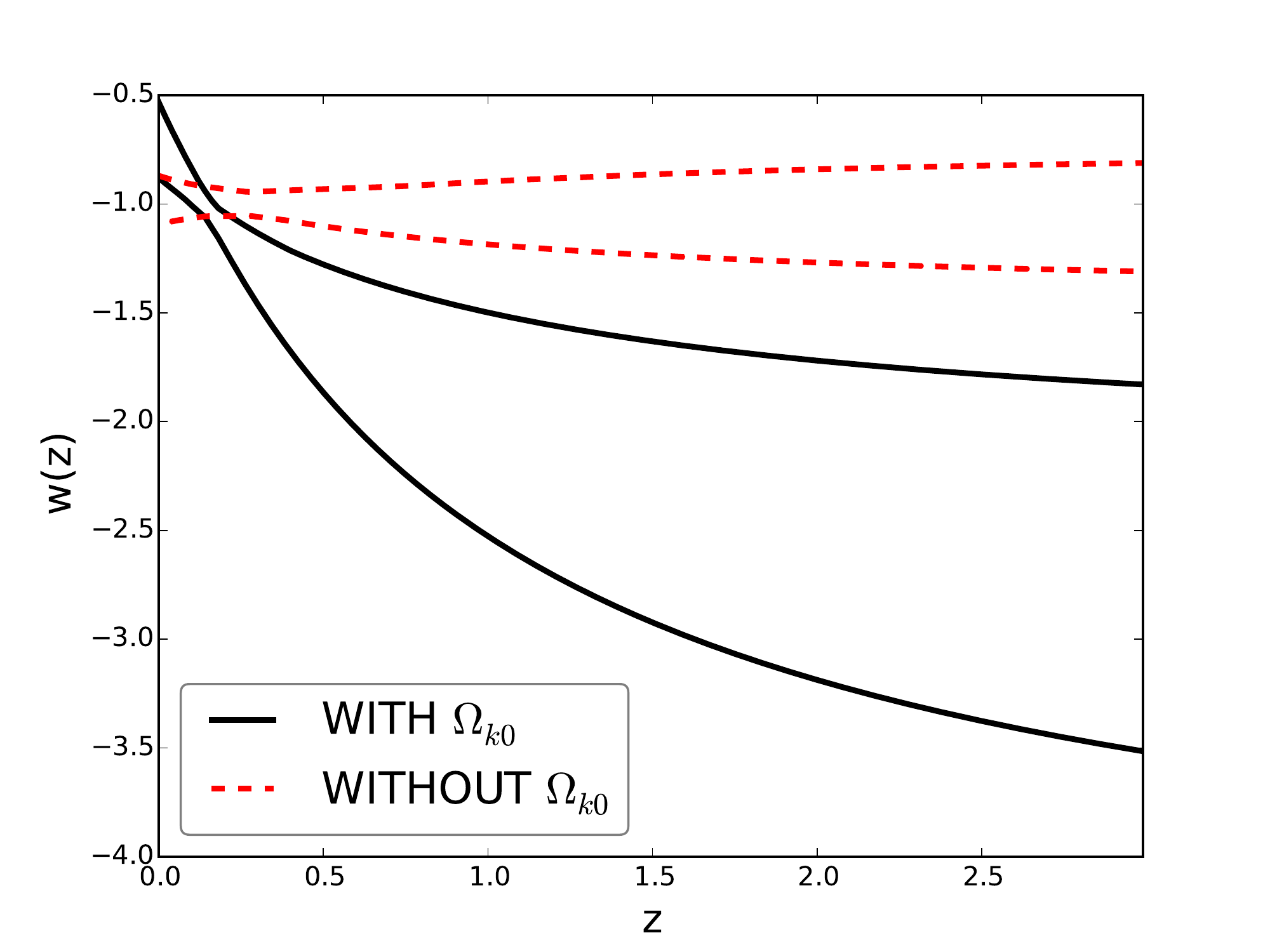}}
  \caption{(color online). The $1\sigma$ and $2\sigma$ probability contours in the $w_0$-$w_a$ plane (left panel)
and the $1\sigma$ confidence regions of $w(z)$ (right panel) for the CPL model.
Black solid lines denote the results obtained in a non-flat Universe,
while red dashed lines denote the results obtained in a flat Universe.
The legends `` WITH $\Omega_{k0}$'' and `` WITHOUT $\Omega_{k0}$'' represent the cases with and without spatial curvature, respectively.
In the left panel, to make a comparison, the fixed point $\{w_0, w_a\} = \{-1,0\}$ of the $\Lambda$CDM model is marked as a cyan star.
In addition, the magenta dashed lines divides the panel into two regions: big rip region and no big rip region.
}
\label{fig:ocpl_eos_flatcompare}
\end{figure*}

In Fig.~\ref{fig:ocpl_eos_flatcompare}, we plot the $1\sigma$ and $2\sigma$ probability contours in the $w_0$-$w_a$ plane (left panel)
and the $1\sigma$ confidence regions of $w(z)$ (right panel) for the CPL model.
From the left panel we see that, the CPL model is consistent with the $\Lambda$CDM model in a flat Universe.
In contrast, the CPL model deviates from the $\Lambda$CDM model at 1$\sigma$ CL in a non-flat Universe;
in other words, in the framework of non-flat Universe, the CPL model will predict a Universe without big rip.
From the right panel we find that, in a flat Universe, $w(z)$ is consistent with $-1$, which corresponds to an eternal CA.
In contrast, in a non-flat Universe, $w(z)$ is a decreasing function of redshift $z$ when $z \rightarrow 0$, which corresponds to a slowing down CA.

\begin{figure*}
  \centering
  \resizebox{0.74\columnwidth}{!}{\includegraphics{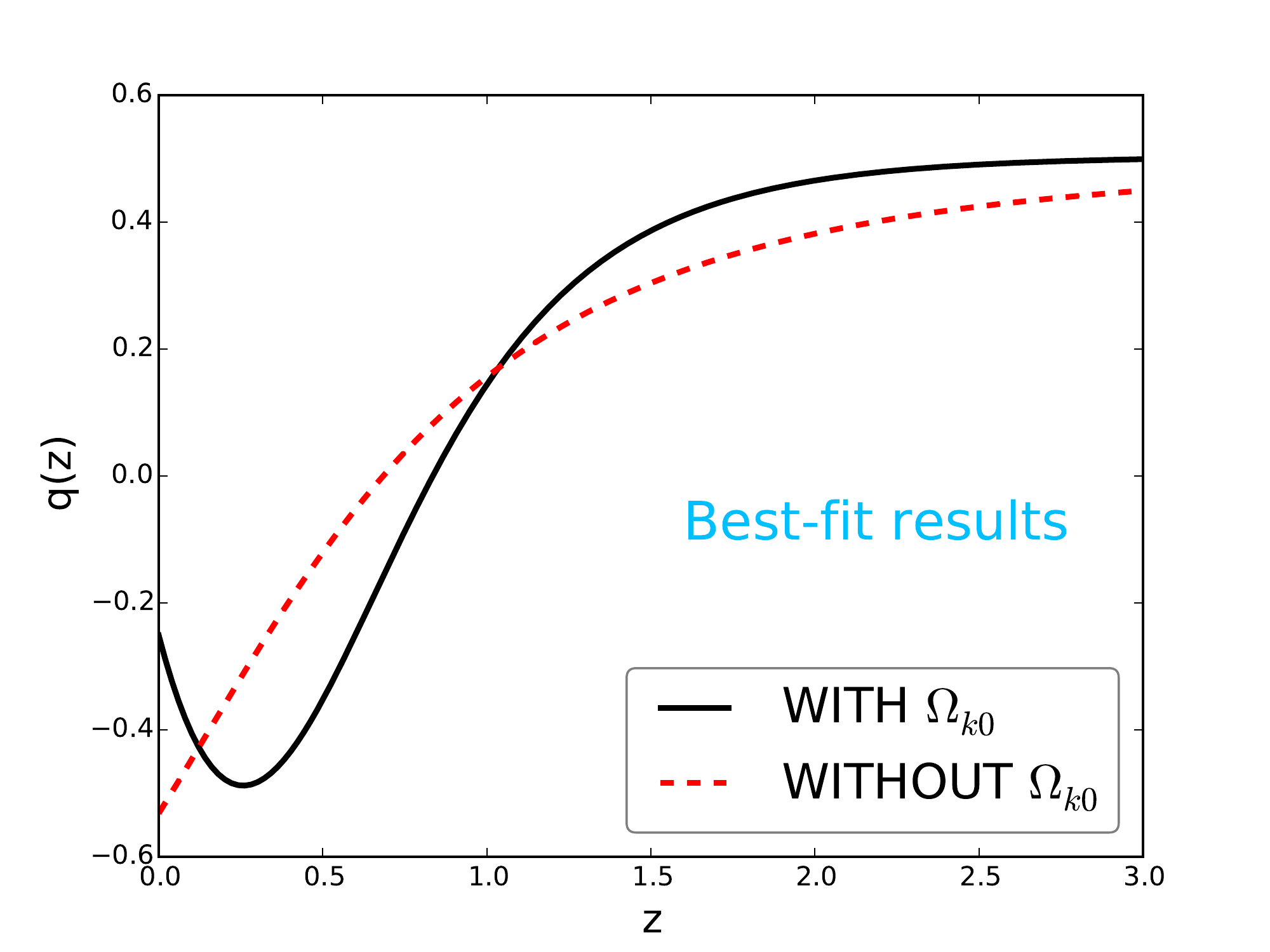}}
  \hspace{0.1\columnwidth}
  \resizebox{0.74\columnwidth}{!}{\includegraphics{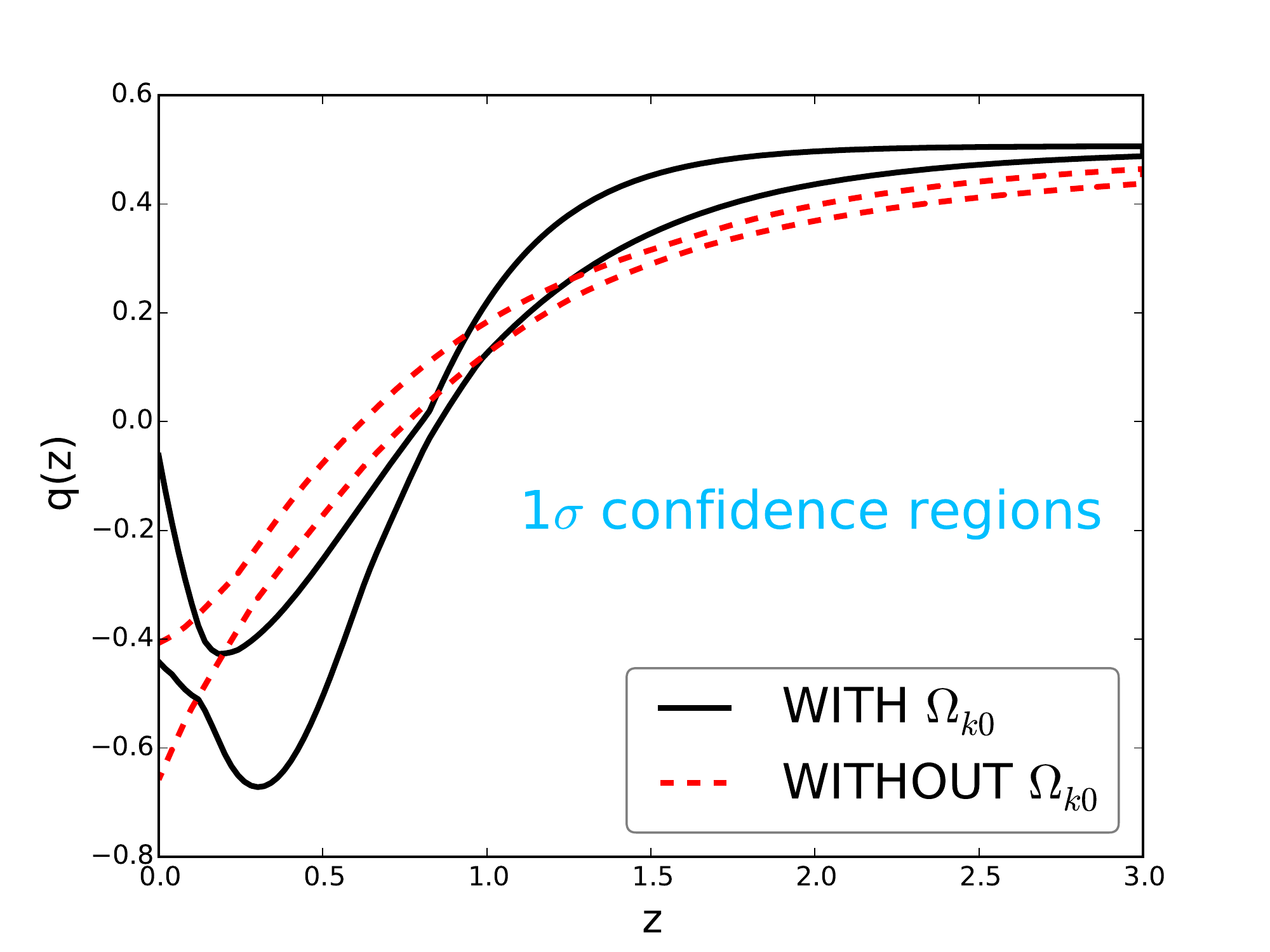}}
  \hspace{0.1\columnwidth}
  \resizebox{0.74\columnwidth}{!}{\includegraphics{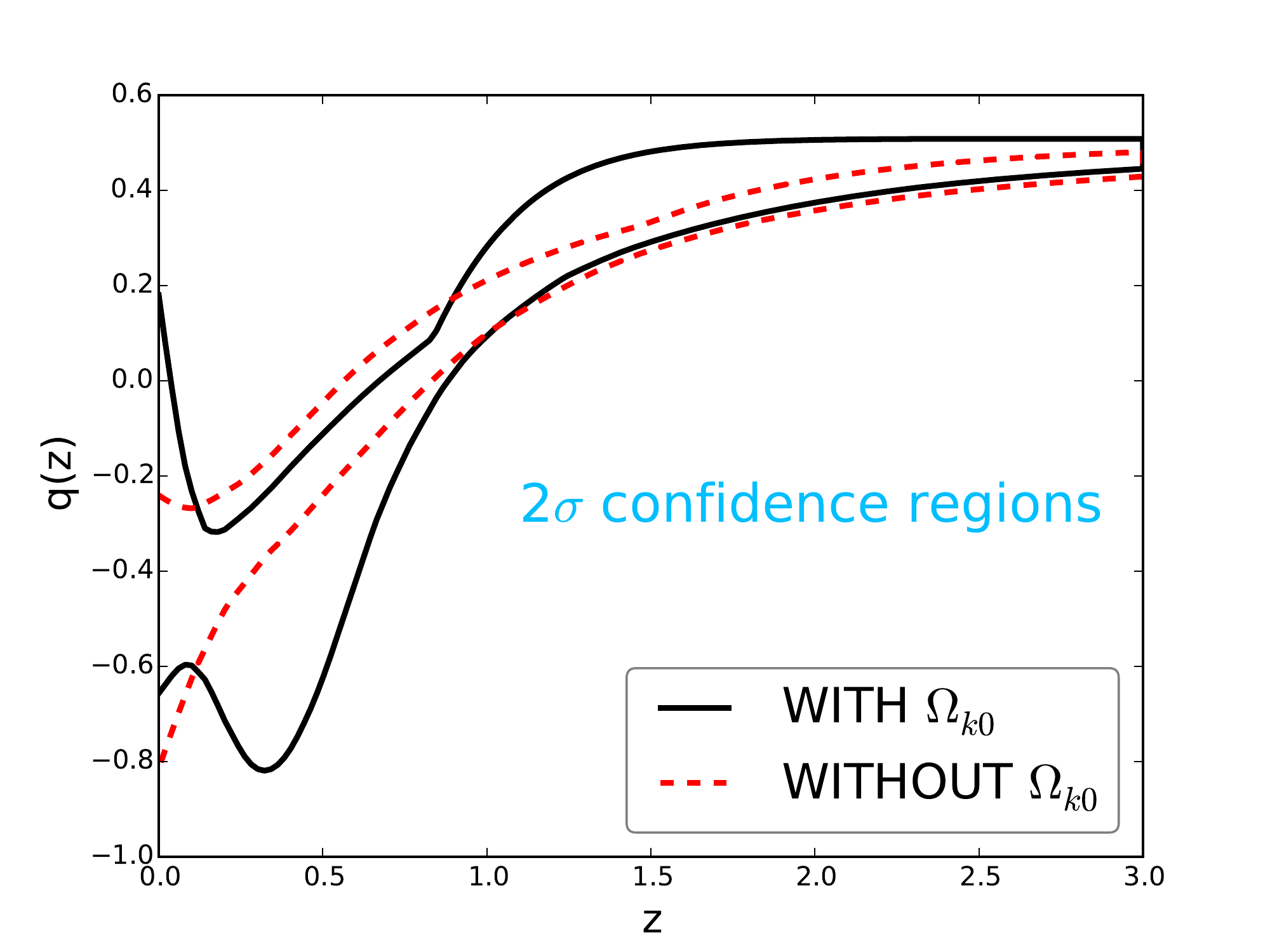}}
  \caption{(color online). The best-fit results (upper left panel), the 1$\sigma$ confidence regions (upper right panel),
and the 2$\sigma$ confidence regions (lower panel) of deceleration parameter $q(z)$ for the CPL model.
Black solid lines denote the results obtained in a non-flat Universe,
while red dashed lines denote the results obtained in a flat Universe.
The legends `` WITH $\Omega_{k0}$'' and `` WITHOUT $\Omega_{k0}$'' represent the cases with and without spatial curvature, respectively.
}
\label{fig:o_qz_flatcompare}
\end{figure*}

Fig.~\ref{fig:o_qz_flatcompare} shows the best-fit results (upper left panel), the 1$\sigma$ confidence regions (upper right panel),
and the 2$\sigma$ confidence regions (lower panel) of deceleration parameter $q(z)$ for the CPL model.
We see that, in a flat Universe, $q(z)$ is always an increasing function of $z$, which corresponds to an eternal CA
In a non-flat Universe, $q(z)$ becomes an decreasing function when $z \rightarrow 0$ at 1$\sigma$ CL,
but is still consistent with an increasing function at 2$\sigma$ CL.
This means that the CPL model in a non-flat Universe favors a slowing down CA at 1$\sigma$ CL.

In conclusion, considering spatial curvature or not will significantly change the evolutionary trajectories of CA.
Using the $\rm SNLS3(linear\ \beta)+BAO(1D)+Planck2015$ data,
a non-flat Universe prefers a slowing down CA at 1$\sigma$ CL,
while a flat Universe favors an eternal CA.

\subsection{Effects of Different observational Data on CA}

Now, we discuss the impacts of different observations, including SNe Ia, BAO, and CMB data, on the evolutionary behavior of CA.
To study the impacts of different observational data, it is necessary to fix the cosmological model in the background first.
So in this subsection, we always use the CPL model in a non-flat Universe.

\subsubsection{Impacts of Different SNe Ia Data}

\begin{figure*}
  \centering
  \resizebox{0.70\columnwidth}{!}{\includegraphics{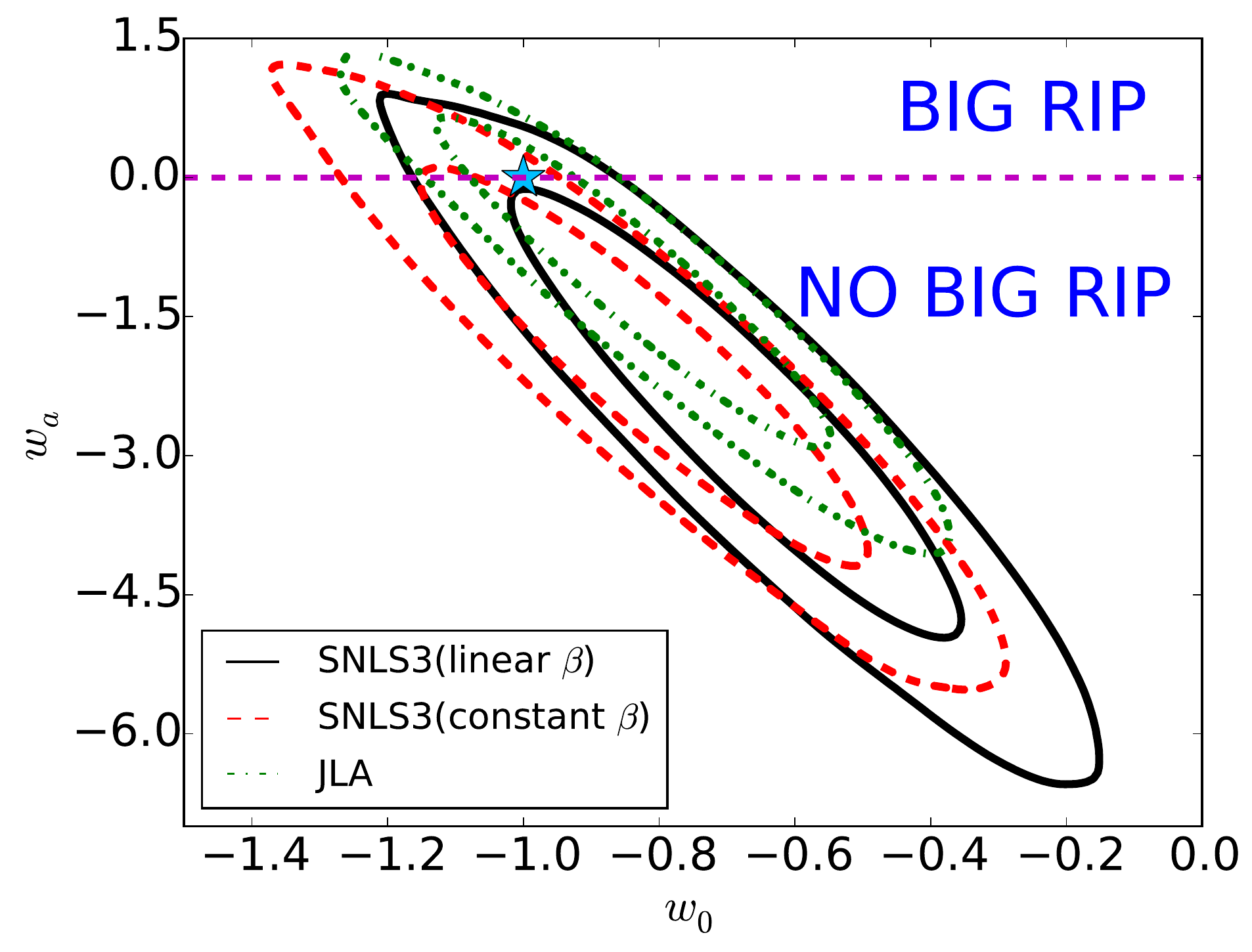}}
  \hspace{0.1\columnwidth}
  \resizebox{0.74\columnwidth}{!}{\includegraphics{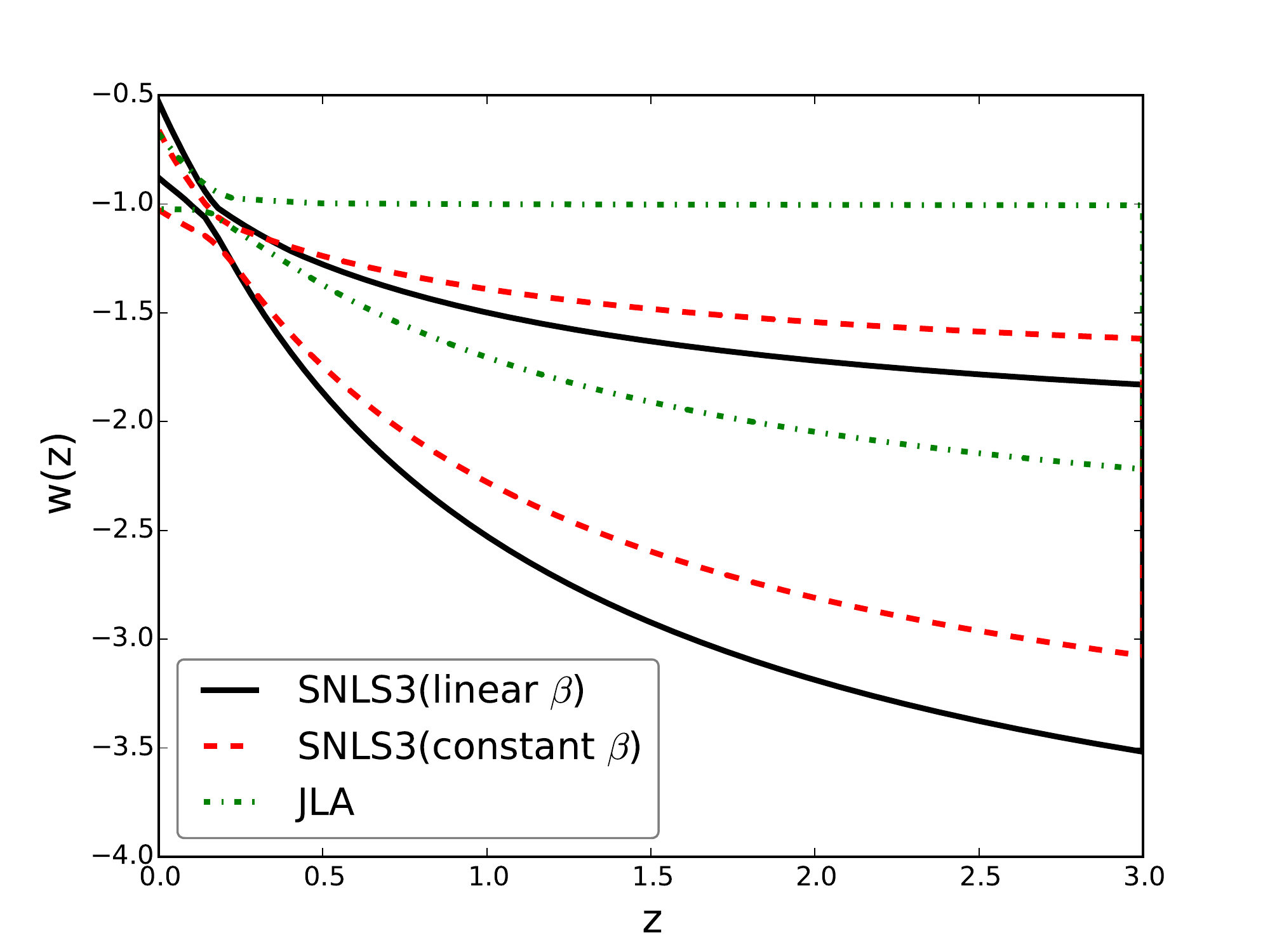}}
  \caption{(color online). The $1\sigma$ and $2\sigma$ probability contours in the $w_0$-$w_a$ plane (left panel)
and the $1\sigma$ confidence regions of $w(z)$ (right panel) given by three kinds of SNe Ia data.
The legends ``SNLS3(linear $\beta$)'', ``SNLS3(constant $\beta$)'' and ``JLA'' represent ``SNLS3(linear $\beta$)+BAO(1D)+Planck2015'',
``SNLS3(constant $\beta$)+BAO(1D)+Planck2015'' and ``JLA+BAO(1D)+Planck2015'' data, respectively.
Black solid lines denote the results given by the ``SNLS3(linear $\beta$)'' data,
red dashed lines denote the results given by the ``SNLS3(constant $\beta$)'' data,
while green dash-dotted lines denote the results given by the ``JLA'' data.
In the left panel, the fixed point $\{w_0, w_a\} = \{-1,0\}$ of the $\Lambda$CDM model is marked as a cyan star.
In addition, the magenta dashed lines divides the panel into two regions: big rip region and no big rip region.
}
\label{fig:ocpl_eos_sncompare}
\end{figure*}

For the observational side,
let us discuss the impacts of different SNe Ia data first.
In Fig.~\ref{fig:ocpl_eos_sncompare}, we plot the $1\sigma$ and $2\sigma$ probability contours in the $w_0$-$w_a$ plane (left panel)
and the $1\sigma$ confidence regions of $w(z)$ (right panel) given by three kinds of SNe Ia data.
From the left panel we see that, the 1$\sigma$ contour given by the ``JLA'' data accommodates the fixed point of the $\Lambda$CDM model;
in contrast, the 1$\sigma$ contours given by the ``SNLS3(linear $\beta$)'' and the ``SNLS3(constant $\beta$)'' data
deviate from the result of the $\Lambda$CDM model.
From the right panel we find that, the $1\sigma$ region of $w(z)$ given by the ``JLA'' data is consistent with $-1$, which corresponds to an eternal CA.
In contrast, the $1\sigma$ regions of $w(z)$ given by the ``SNLS3(linear $\beta$)'' and the ``SNLS3(constant $\beta$)'' data
are decreasing functions of $z$ when $z \rightarrow 0$, which corresponds to a slowing down CA.
In addition, the results given by the ``SNLS3(constant $\beta$)'' data are a little closer to a cosmological constant than
the results of the ``SNLS3(linear $\beta$)'' data.

\begin{figure*}
  \centering
  \resizebox{0.74\columnwidth}{!}{\includegraphics{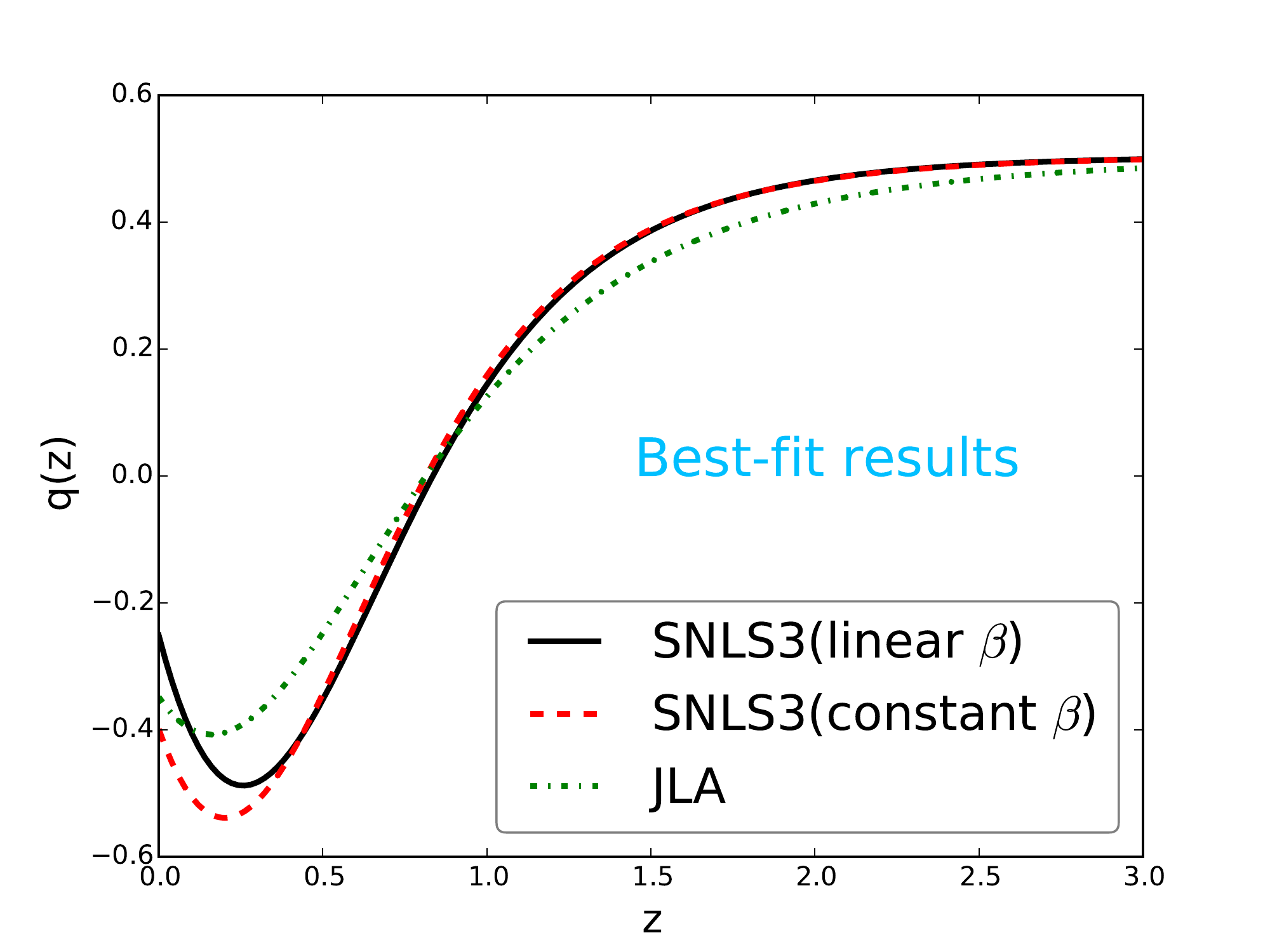}}
  \hspace{0.1\columnwidth}
  \resizebox{0.74\columnwidth}{!}{\includegraphics{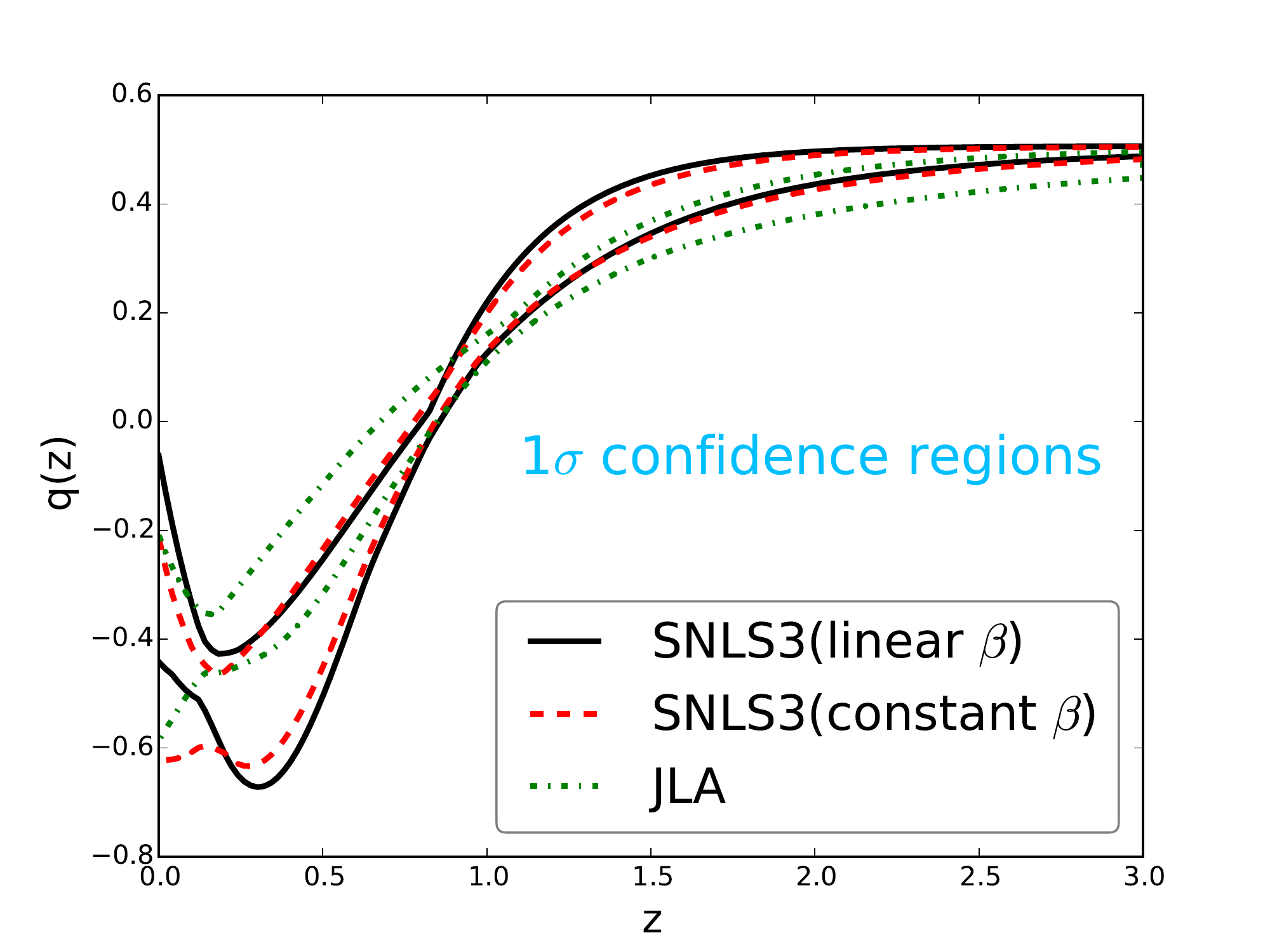}}
  \hspace{0.1\columnwidth}
  \resizebox{0.74\columnwidth}{!}{\includegraphics{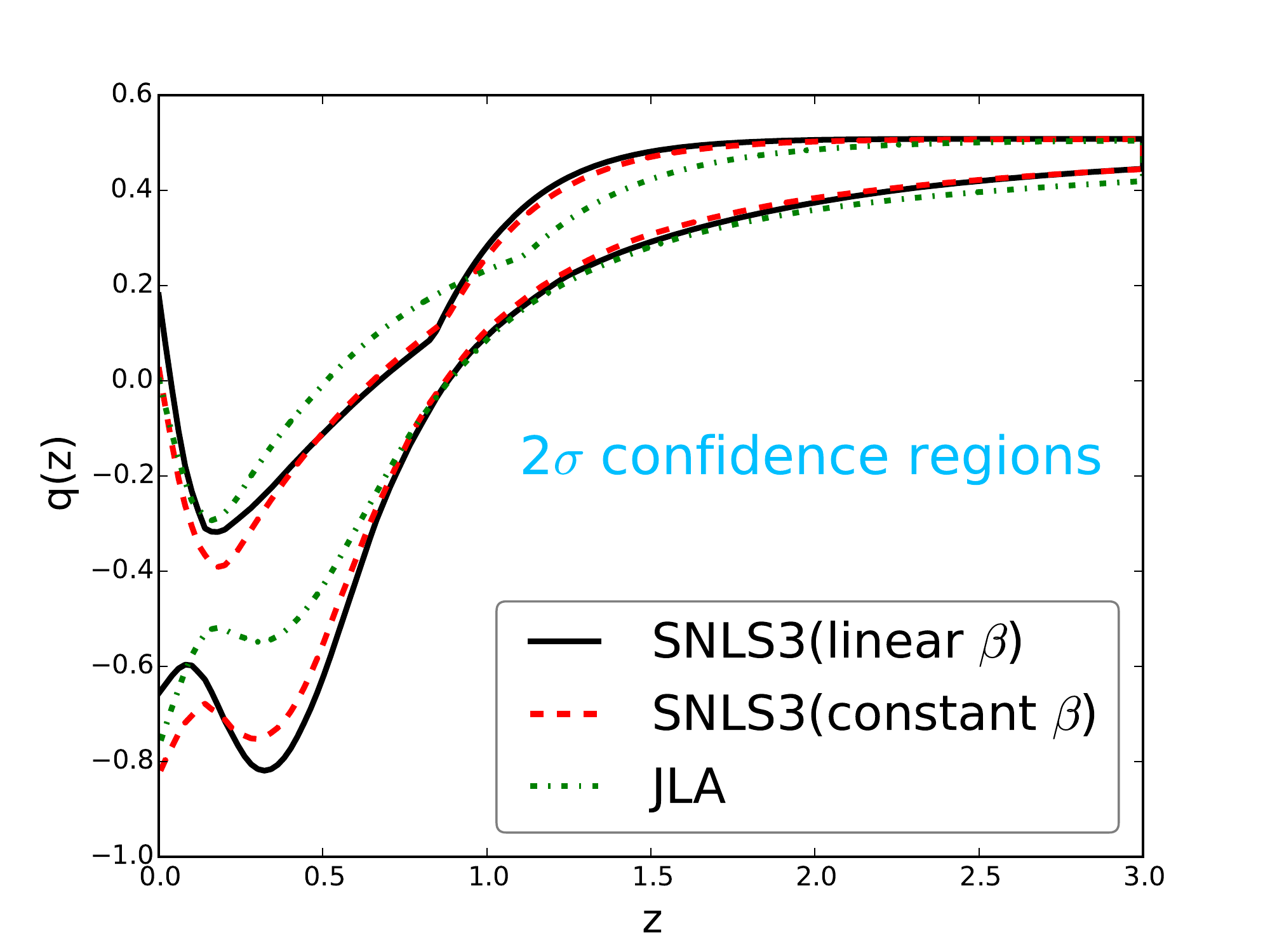}}
  \caption{(color online). The best-fit results (upper left panel), the 1$\sigma$ confidence regions (upper right panel)
and the 2$\sigma$ confidence regions (lower panel) of deceleration parameter $q(z)$ given by three kinds of SNe Ia data.
The legend ``SNLS3(linear $\beta$)'', ``SNLS3(constant $\beta$)'' and ``JLA''
represent ``SNLS3(linear $\beta$)+BAO(1D)+Planck2015'', ``SNLS3(constant $\beta$)+BAO(1D)+Planck2015'' and ``JLA+BAO(1D)+Planck2015'' data, respectively.
Black solid lines denote the results given by the ``SNLS3(linear $\beta$)'' data,
red dashed lines denote the results given by the ``SNLS3(constant $\beta$)'' data,
while green dash-dotted lines denote the results given by the ``JLA'' data.
}
\label{fig:o_qz_sncompare}
\end{figure*}

In Fig.~\ref{fig:o_qz_sncompare}, we plot the best-fit results (upper left panel), the 1$\sigma$ confidence regions (upper right panel)
and the 2$\sigma$ confidence regions (lower panel) of deceleration parameter $q(z)$ given by three kinds of SNe Ia data.
We see that,
both the best-fit results of $q(z)$ given by the ``JLA'', the ``SNLS3(linear $\beta$)''
and the ``SNLS3(constant $\beta$)'' data correspond to a slowing down CA.
At 1$\sigma$ CL, the results given by the ``JLA'' data are consistent with an eternal CA,
while the ``SNLS3(linear $\beta$)'' and the ``SNLS3(constant $\beta$)'' data still favor a slowing down CA.
At 2$\sigma$ CL, both the ``JLA'', the ``SNLS3(linear $\beta$)''
and the ``SNLS3(constant $\beta$)'' data are consistent with an eternal CA.
This means that the SNLS3 data favor a slowing down CA at 1$\sigma$ CL,
while the ``JLA'' data prefer an an eternal CA at 1$\sigma$ CL.

In conclusion, the use of SNe Ia data has significant impacts on the evolutionary behavior of CA:
the SNLS3 datasets favor a slowing down CA at 1$\sigma$ CL,
while the JLA samples prefer an eternal CA.
Without considering spatial curvature,
Magana et al. found that
the Union2.1 data~\citep{Suzuki2012} prefer an eternal CA;
in contrast, the Constitution~\citep{Hicken2009}, the Union2~\citep{Amanullah2010} and the LOSS-Union~\citep{Ganeshalingam2013} SNe Ia data favor a slowing down CA at 1$\sigma$ CL~\citep{Magana2014}.
Therefore, our study verifies the conclusion of \cite{Magana2014}.

\subsubsection{Impacts of Different BAO Data}

\begin{figure*}
  \centering
  \resizebox{0.70\columnwidth}{!}{\includegraphics{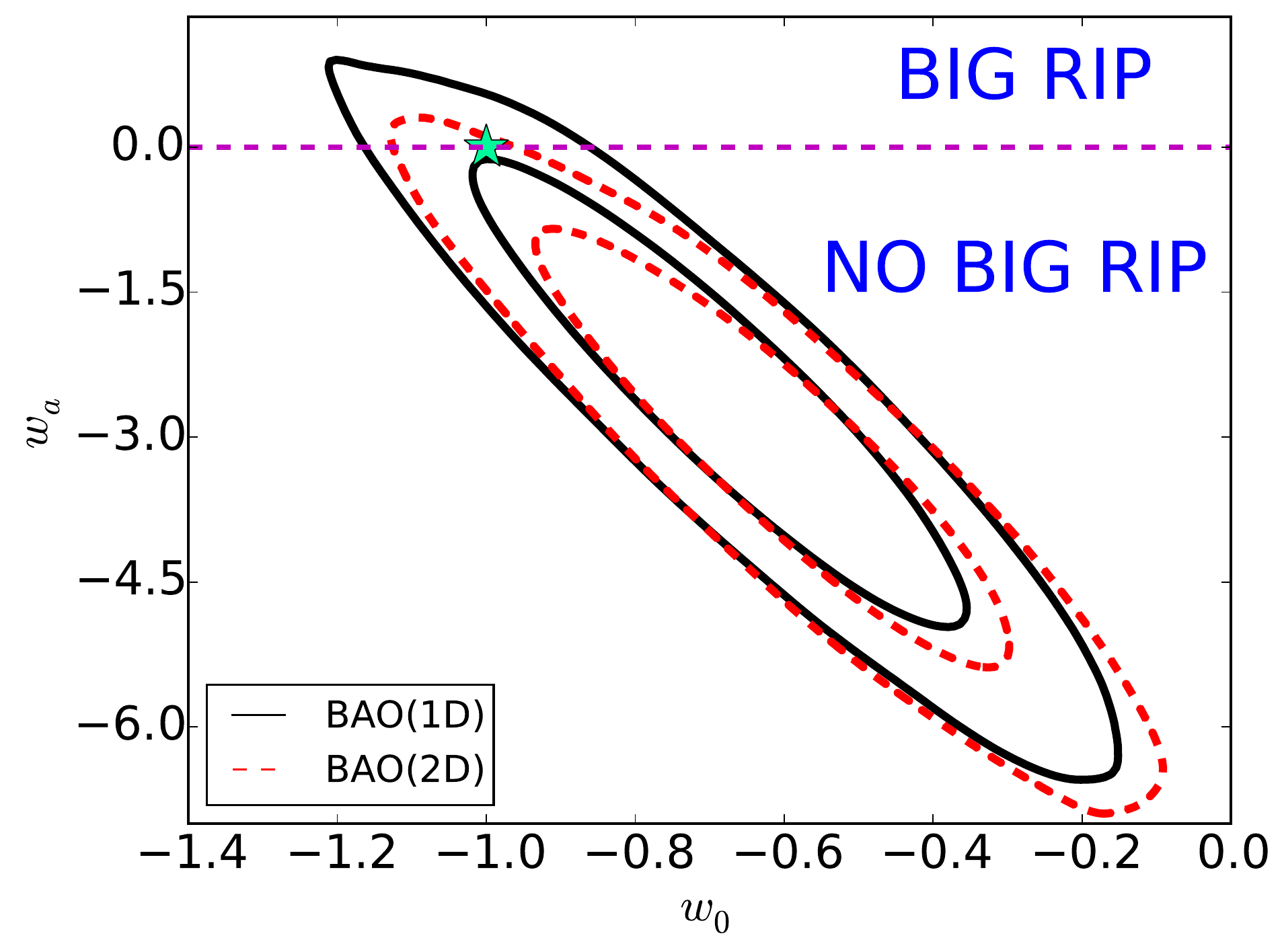}}
  \hspace{0.1\columnwidth}
  \resizebox{0.74\columnwidth}{!}{\includegraphics{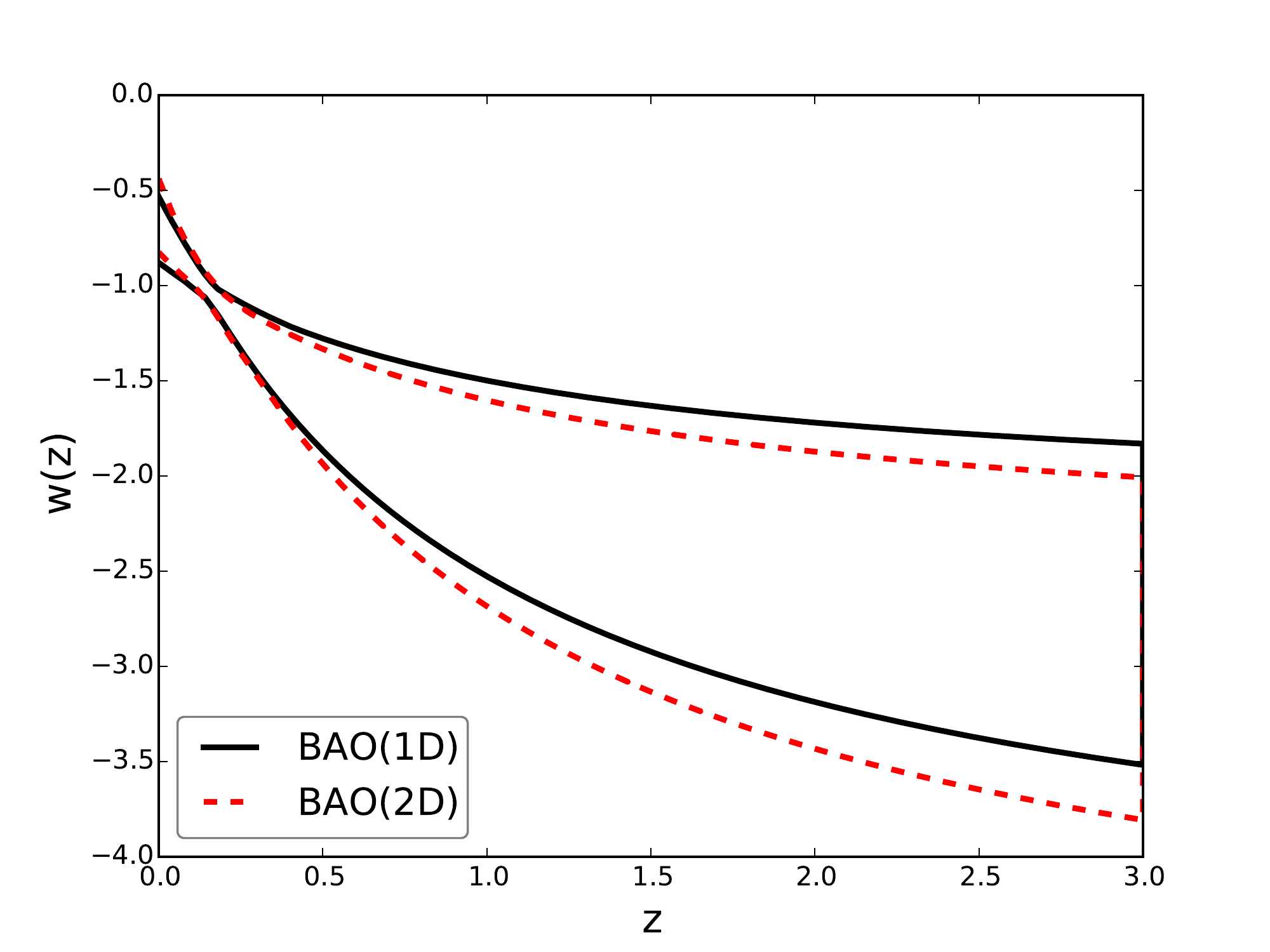}}
  \caption{(color online). The $1\sigma$ and $2\sigma$ probability contours in the $w_0$-$w_a$ plane (left panel)
and the $1\sigma$ confidence regions of $w(z)$ (right panel) given by two kinds of BAO data.
The legend ``BAO(1D)'' and ``BAO(2D)'' represent the ``SNLS3(linear $\beta$)+BAO(1D)+Planck2015''
and ``SNLS3(linear $\beta$)+BAO(2D)+Planck2015'' data, respectively.
Black solid lines denote the results given by the ``BAO(1D)'' data,
while red dashed lines denote the results given by the ``BAO(2D)'' data.
In the left panel, the fixed point $\{w_0, w_a\} = \{-1,0\}$ of the $\Lambda$CDM model is marked as a cyan star.
In addition, the magenta dashed lines divides the panel into two regions: big rip region and no big rip region.
}
\label{fig:ocpl_eos_baocompare}
\end{figure*}

Then, let us turn to the impacts of different BAO data.
In Fig.~\ref{fig:ocpl_eos_baocompare}, we plot the $1\sigma$ and $2\sigma$ probability contours in the $w_0$-$w_a$ plane (left panel)
and the $1\sigma$ confidence regions of $w(z)$ (right panel) given by two kinds of BAO data.
From the left panel we see that, the 1$\sigma$ contours given by the ``BAO(1D)'' and the ``BAO(2D)'' data
cannot accommodate the fixed point of the $\Lambda$CDM model.
From the right panel we find that, the $1\sigma$ regions of $w(z)$ given by the ``BAO(1D)'' and the ``BAO(2D)'' data
are decreasing functions of $z$ when $z \rightarrow 0$, which corresponds to a slowing down CA.
In addition, the ``BAO(1D)'' data give a little larger $q(z)$, which is a little closer to the case of the $\Lambda$CDM model.

\begin{figure*}
  \centering
  \resizebox{0.74\columnwidth}{!}{\includegraphics{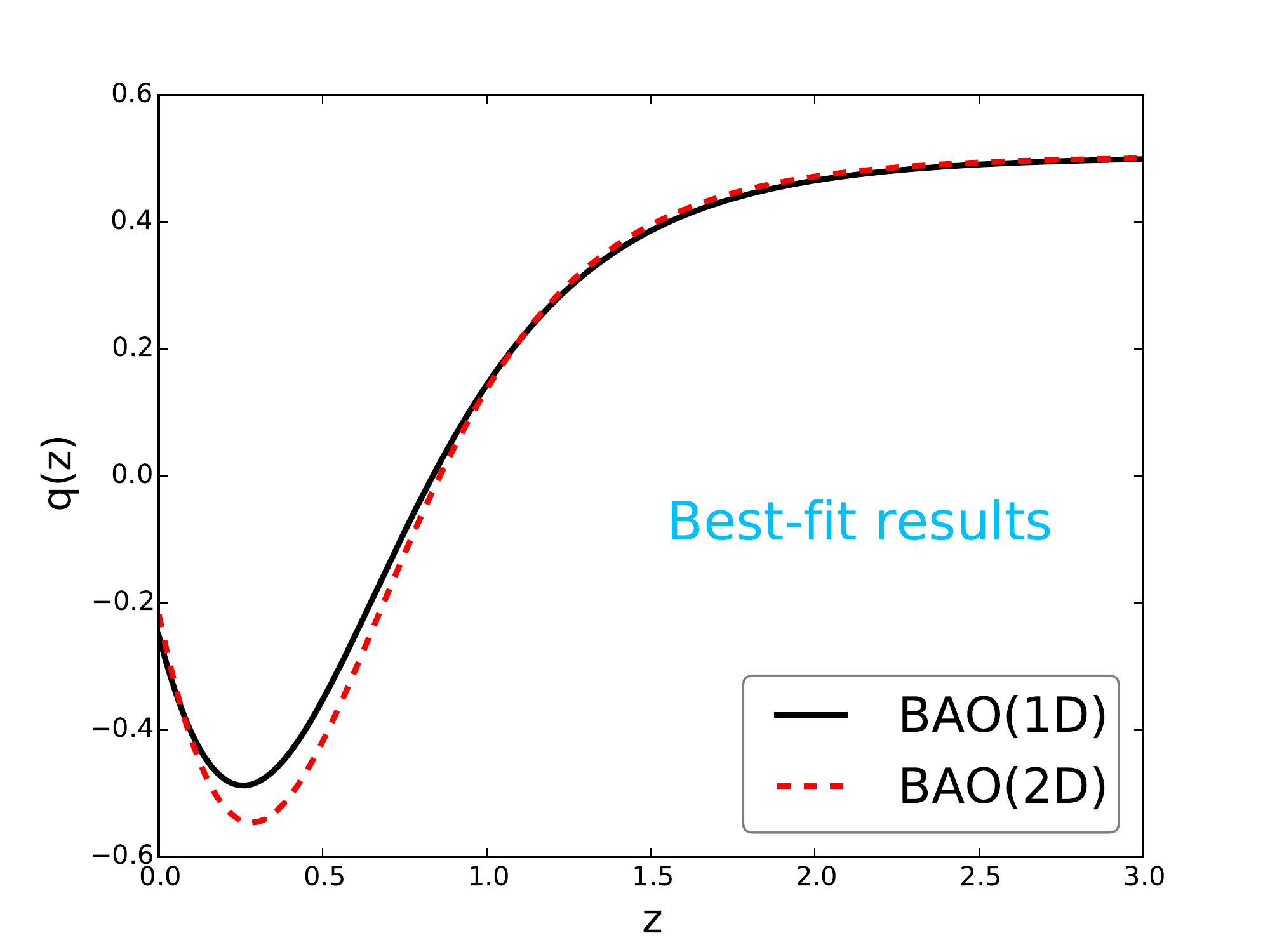}}
  \hspace{0.1\columnwidth}
  \resizebox{0.74\columnwidth}{!}{\includegraphics{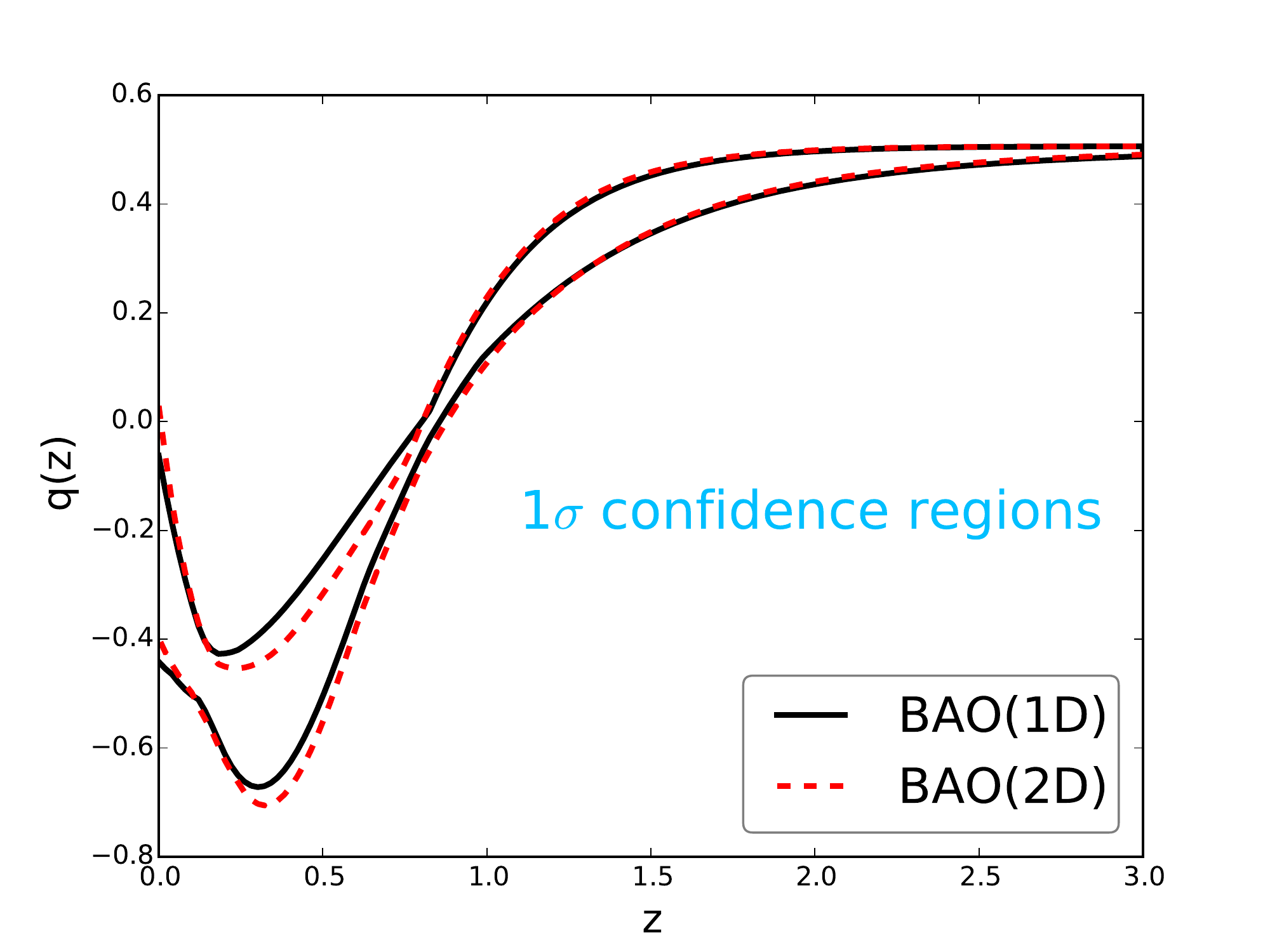}}
  \hspace{0.1\columnwidth}
  \resizebox{0.74\columnwidth}{!}{\includegraphics{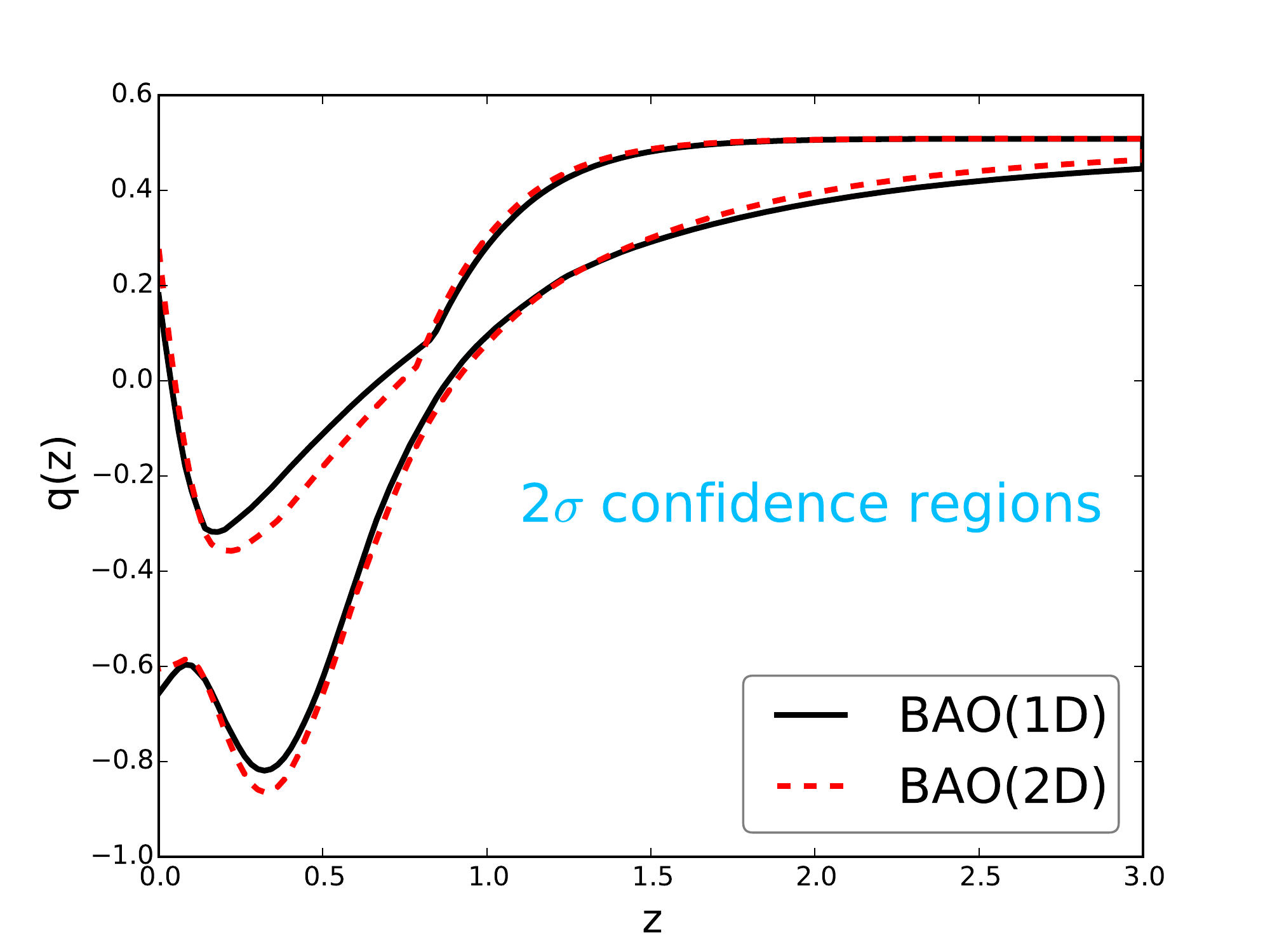}}
    \caption{(color online).  The best-fit results (upper left panel), the 1$\sigma$ confidence regions (upper right panel)
and the 2$\sigma$ confidence regions (lower panel) of deceleration parameter $q(z)$ given by two kinds of BAO data.
The legend ``BAO(1D)'' and ``BAO(2D)'' represent the ``SNLS3(linear $\beta$)+BAO(1D)+Planck2015''
and ``SNLS3(linear $\beta$)+BAO(2D)+Planck2015'' data, respectively.
Black solid lines denote the results given by the ``BAO(1D)'' data,
while red dashed lines denote the results given by the ``BAO(2D)'' data.
}
\label{fig:o_qz_baocompare}
\end{figure*}

In Fig.~\ref{fig:o_qz_baocompare}, we plot the best-fit results (upper left panel), the 1$\sigma$ confidence regions (upper right panel)
and the 2$\sigma$ confidence regions (lower panel) of deceleration parameter $q(z)$ given by two kinds of BAO data.
We see that, for both the cases of ``BAO(1D)'' and ``BAO(2D)'',
$q(z)$ undergoes a transition at $z\simeq0.25$, an then becomes an decreasing function when $z \rightarrow 0$ at 1$\sigma$ CL;
In addition, this feature disappears at 2$\sigma$ CL,
This shows that both the ``BAO(1D)'' and the ``BAO(2D)'' data favor a slowing down CA at 1$\sigma$ CL.
Moreover, we find that the differences between the evolutionary trajectories of $q(z)$ given by the ``BAO(1D)'' and the ``BAO(2D)'' data are very small.

In conclusion, the effects of different BAO data on CA are negligible.

\subsubsection{Impacts of Different CMB Data}

Finally, let us discuss the impacts of different CMB data.
To do this, we make two comparisons for the CMB data: 1. a comparison among different CMB distance prior data;
2. a comparison between the CMB distance prior data and the full CMB data.
It must be emphasized that, in order to get better visual effects about these comparisons,
in this subsubsection we make use of the JLA SNe Ia samples, instead of the  ``SNLS3(linear $\beta$) data.

\begin{figure*}
  \centering
  \resizebox{0.70\columnwidth}{!}{\includegraphics{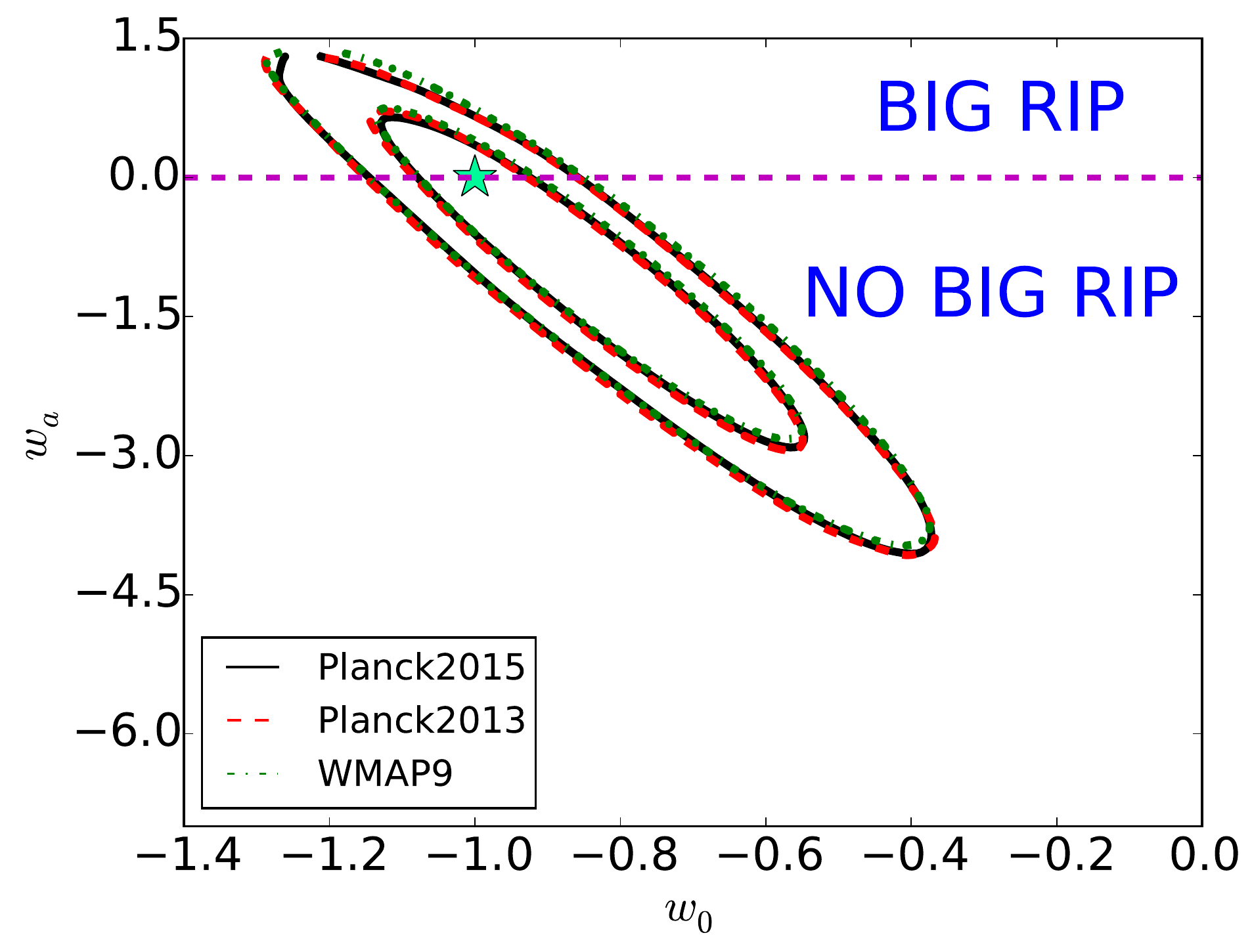}}
  \hspace{0.1\columnwidth}
  \resizebox{0.70\columnwidth}{!}{\includegraphics{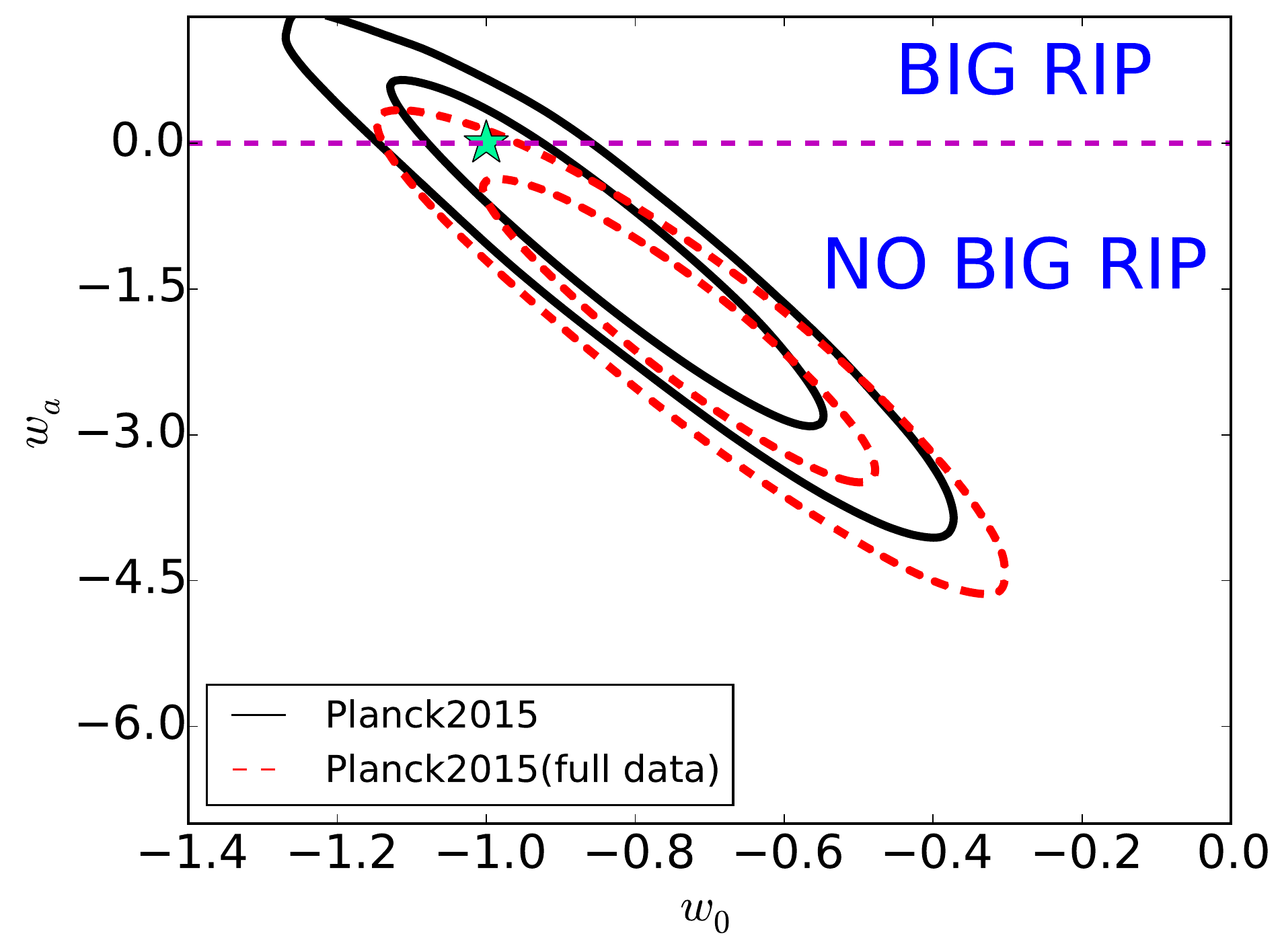}}
  \hspace{0.1\columnwidth}
  \resizebox{0.74\columnwidth}{!}{\includegraphics{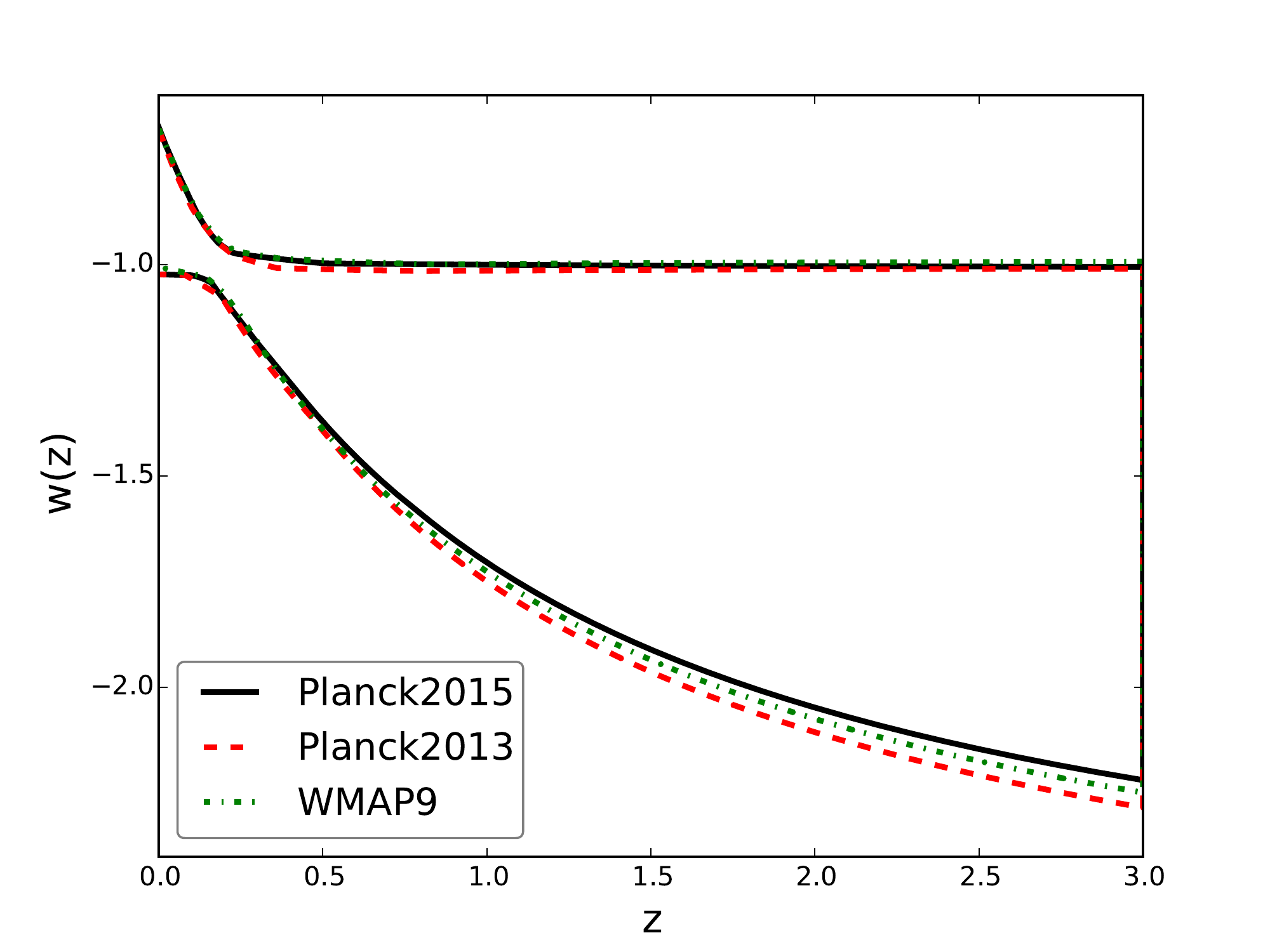}}
  \hspace{0.1\columnwidth}
  \resizebox{0.74\columnwidth}{!}{\includegraphics{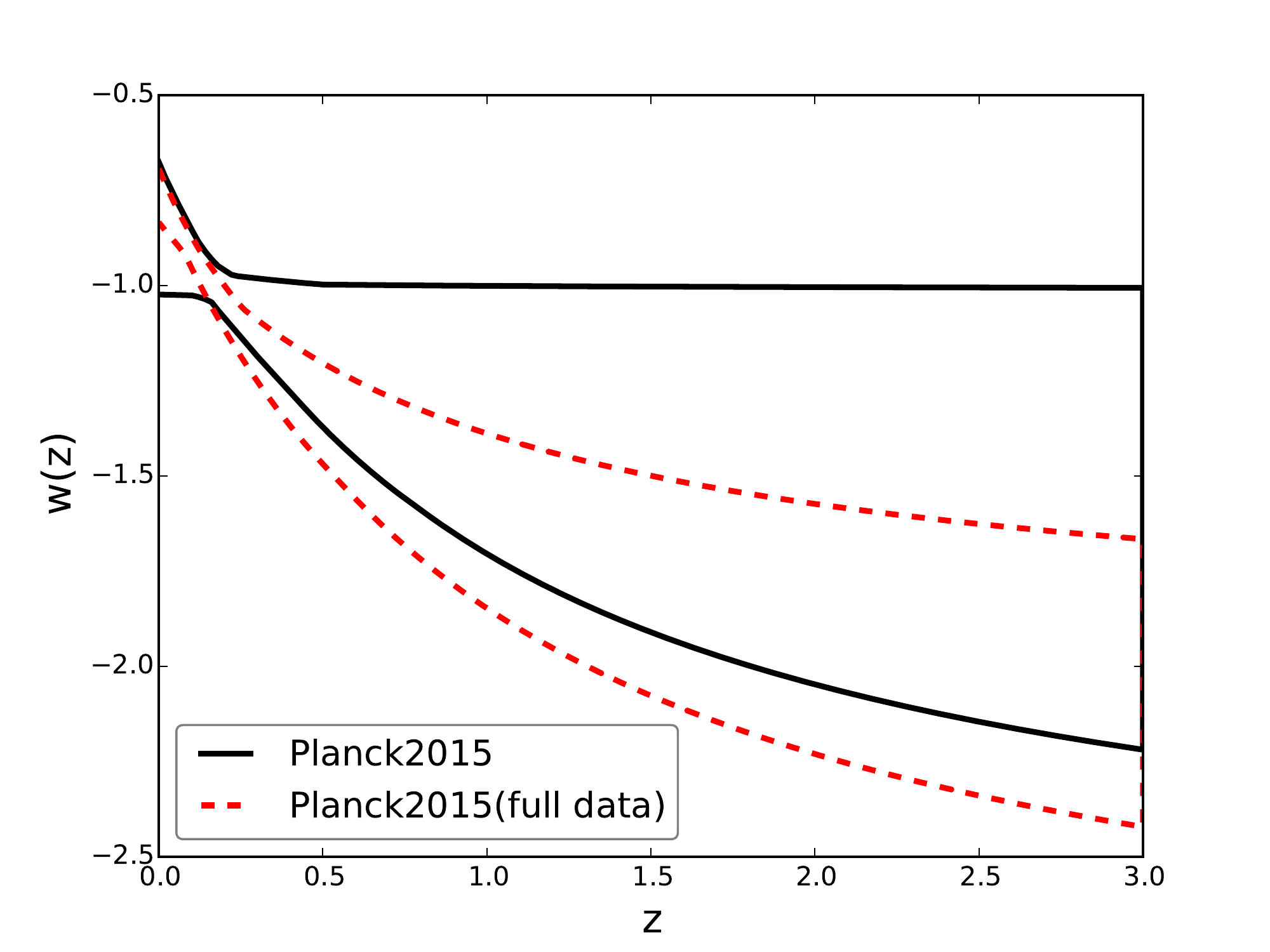}}
  \caption{(color online). The $1\sigma$ and $2\sigma$ probability contours in the $w_0$-$w_a$ plane
given by three kinds of CMB distance prior data (upper left panel) and two kinds of Planck2015 data (upper right panel),
as well as the $1\sigma$ confidence regions of $w(z)$
given by three kinds of CMB distance prior data (lower left panel) and two kinds of Planck2015 data (lower right panel).
The legends ``Planck2015'', ``Planck2013'', ``WMAP9'', and ``Planck2015(full data)''
represent the ``JLA+BAO(1D)+Planck2015'', ``JLA+BAO(1D)+Planck2013'', ``JLA+BAO(1D)+WMAP9'', and ``JLA+BAO(1D)+Planck2015(full data)'' data, respectively.
In the left two panels,
black solid lines denote the results given by the ``Planck2015'' data,
red dashed lines denote the results given by the ``Planck2013'' data,
while green dash-dotted lines denote the results given by the ``WMAP9'' data.
In the right two panels,
black solid lines denote the results given by the ``Planck2015'' data,
while red dashed lines denote the results given by the ``Planck2015(full data)'' data.
}
\label{fig:ocpl_eos_cmbcompare}
\end{figure*}

In Fig.~\ref{fig:ocpl_eos_cmbcompare}, we plot the $1\sigma$ and $2\sigma$ probability contours in the $w_0$-$w_a$ plane
given by three kinds of CMB distance prior data (upper left panel) and by two kinds of Planck2015 data (upper right panel),
as well as the $1\sigma$ confidence regions of $w(z)$
given by three kinds of CMB distance prior data (lower left panel) and by two kinds of Planck2015 data (lower right panel).
As mentioned above, to get better visual effects about the comparisons of various CMB data,
the JLA data are used in the analysis, and this is why the 1$\sigma$ contours
given by the ``Planck2015'', the ``Planck2013'', and the ``WMAP9'' data
can accommodate the fixed point of the $\Lambda$CDM model (see Fig.~\ref{fig:ocpl_eos_sncompare} for details).
From the left two panels we see that,
both the probability contours of $\{w_0,w_a\}$ and the confidence regions of $w(z)$ given by the three kinds of CMB distance prior data are almost overlap;
this means that the impacts of different CMB distance prior data on CA are negligible.
In contrast, from the right two panels we find that, the 1$\sigma$ contours
given by the ``Planck2015(full data)'' data cannot accommodate the fixed point of the $\Lambda$CDM model;
correspondingly, the $1\sigma$ region of $w(z)$ given by the ``Planck2015(full data)'' data
is a decreasing function of $z$ when $z \rightarrow 0$, which corresponds to a slowing down CA.
In addition, the differences between the ``Planck2015(full data)'' results and the ``Planck2015'' results are quite large.

\begin{figure*}
  \centering
  \resizebox{0.74\columnwidth}{!}{\includegraphics{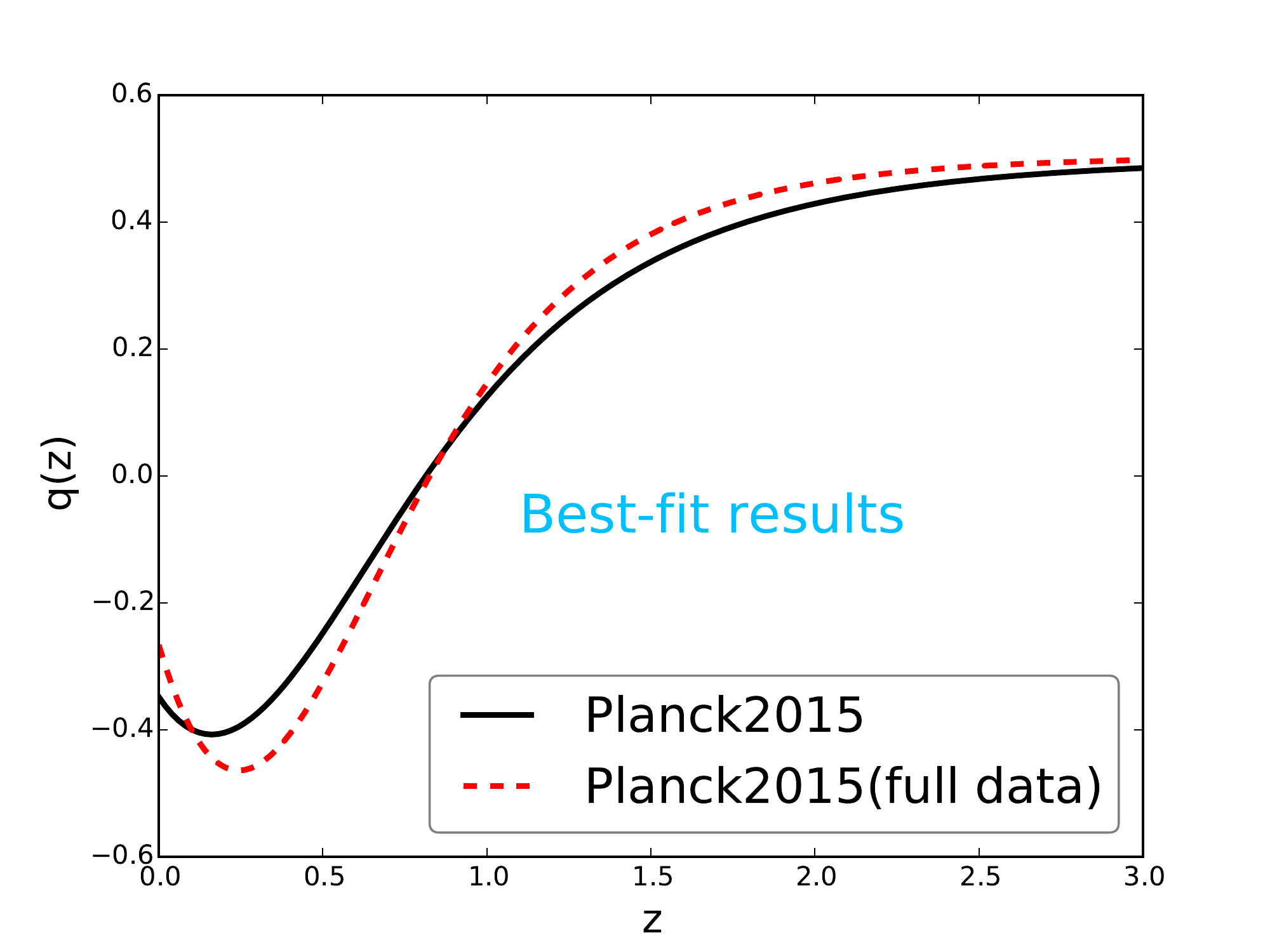}}
  \hspace{0.1\columnwidth}
  \resizebox{0.74\columnwidth}{!}{\includegraphics{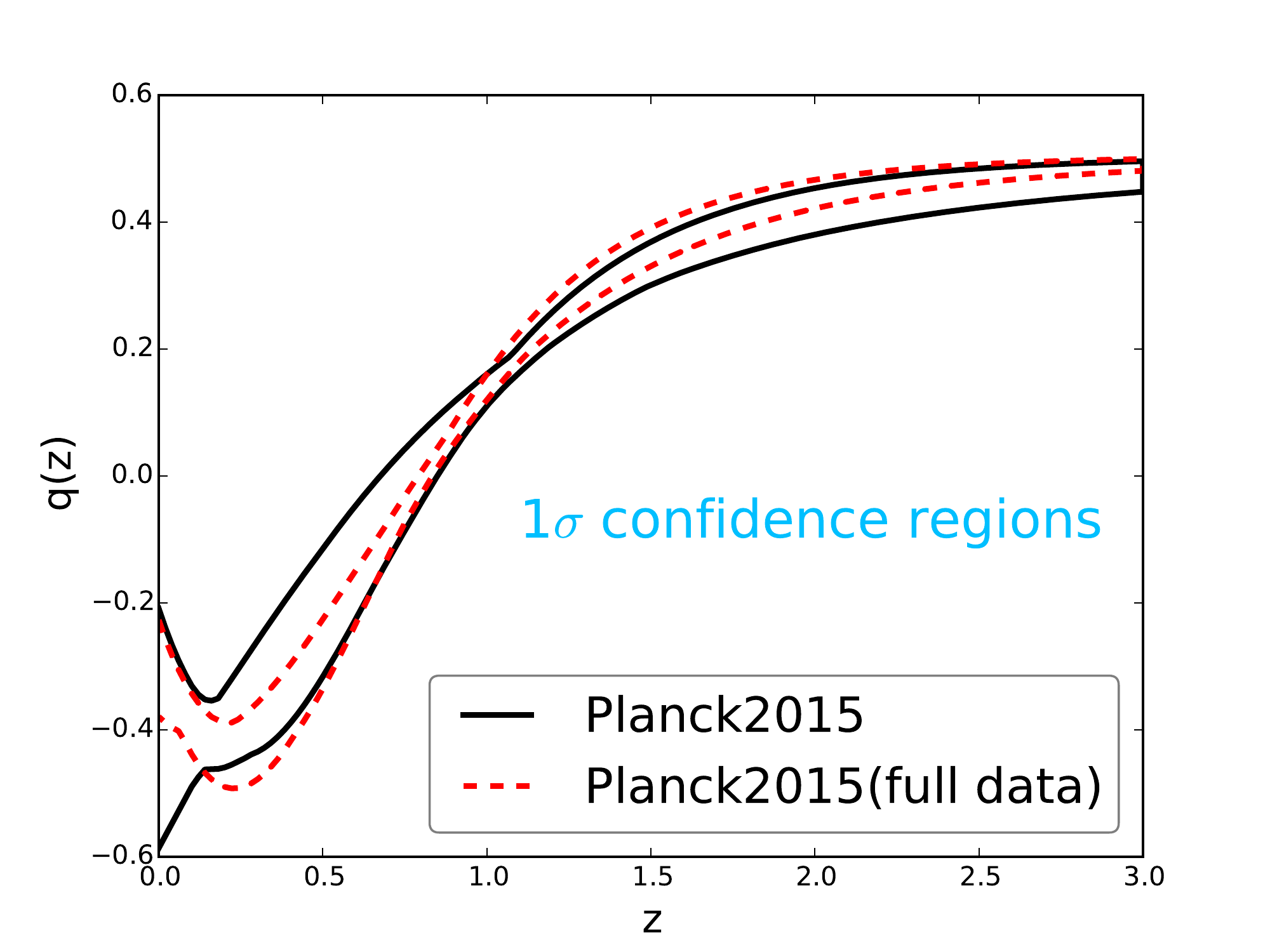}}
  \hspace{0.1\columnwidth}
  \resizebox{0.74\columnwidth}{!}{\includegraphics{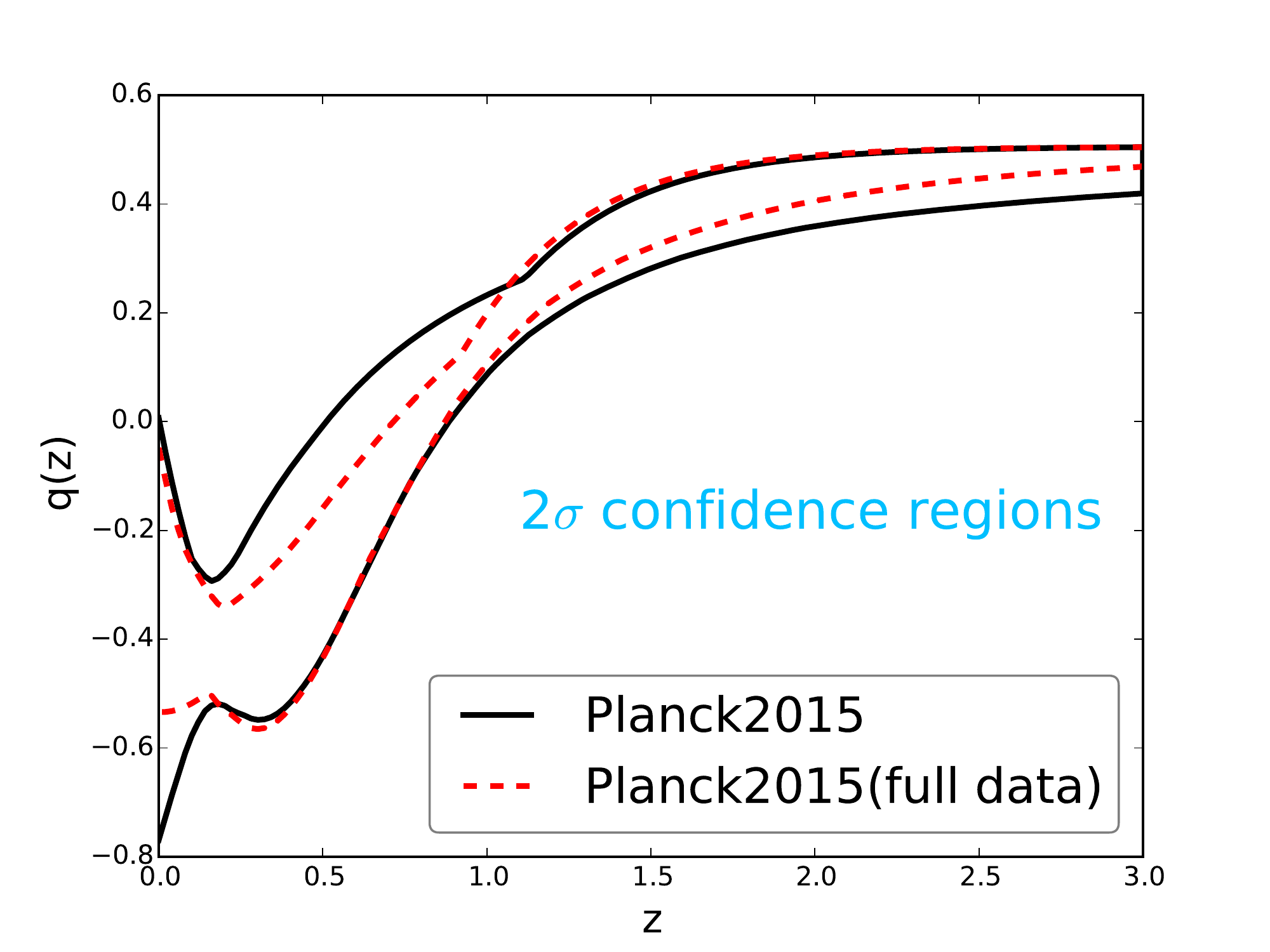}}
  \caption{(color online). The best-fit results (upper left panel), the 1$\sigma$ confidence regions (upper right panel)
and the 2$\sigma$ confidence regions (lower panel) of deceleration parameter $q(z)$
given by  the ``Planck2015'' and the ``Planck2015(full data)'' data.
The legends ``Planck2015'' and ``Planck2015(full data)''
represent the ``JLA+BAO(1D)+Planck2015'' and ``JLA+BAO(1D)+Planck2015(full data)'' data, respectively.
Black solid lines denote the results given by the ``Planck2015'' data,
while red dashed lines denote the results given by the ``Planck2015(full data)'' data.
}
\label{fig:o_qz_cmbcompare}
\end{figure*}

In Fig.~\ref{fig:o_qz_cmbcompare}, we plot the best-fit results (upper left panel), the 1$\sigma$ confidence regions (upper right panel)
and the 2$\sigma$ confidence regions (lower panel) of deceleration parameter $q(z)$
given by the ``Planck2015'' and the ``Planck2015(full data)'' data.
\footnote{Since the differences among the evolutionary trajectories of $q(z)$ given by three kinds of CMB distance prior data are very very small,
the corresponding results are not discussed here.}
Notice that the JLA data are used in the analysis for better visual effects.
From this figure we see that,
both the best-fit results of $q(z)$ given by the ``Planck2015'' and the ``Planck2015(full data)'' data correspond to a slowing down CA.
At 1$\sigma$ CL, the results given by the ``Planck2015'' data are consistent with an eternal CA,
while the ``Planck2015(full data)'' data still favor a slowing down CA.
At 2$\sigma$ CL, both the ``Planck2015'' and the ``Planck2015(full data)'' data are consistent with an eternal CA.
This means that after adopting the JLA data,
the ``Planck2015'' data favor an an eternal CA at 1$\sigma$ CL,
while the ``Planck2015(full data)'' data prefer a slowing down CA at 1$\sigma$ CL.

In conclusion, compared with CMB distance prior data, full CMB data more favor a slowing down CA.
As far as we know, the effects of different BAO and CMB data on CA have not been studied in the past.

\section{Summary and Discussion}
\label{sec:conclusion}

As is well known, in recent 15 years, CA has become one of the biggest puzzles in modern cosmology.
Currently, almost all the mainstream cosmological models predict an eternal CA.
But in~\citep{Shafieloo2009}, Shafieloo, Sanhi, and Starobinsky proposed the possibility that the current CA is slowing down.
Since this extremely counterintuitive phenomenon cannot be accommodated in almost all the mainstream models,
this topic has attracted a lot of interests~\citep{Huang2009, Gong2010, LiWuYu2011, Lin2013, Cardenas2012, Cardenas2013, Cardenas2014} in recent years.

Previous studies on this issue mainly focus on the effects of different SNe Ia datasets;
in addition, most of these works only consider the CPL model in a flat Universe.
A relatively comprehensive study was given in~\citep{Magana2014},
where Magana et al. investigated the evolutionary trajectories of deceleration parameter $q(z)$
by using five DE parametrization models and four SNe Ia datasets.
However, some other factors, including the impacts of spatial curvature, as well as the effects of different BAO and CMB data, are not considered in Ref.~\citep{Magana2014}.

In the present work, by taking into account all the factors mentioned above,
we perform a comprehensive and systematic investigation on the slowing down of CA from both the theoretical and the observational sides.
For the theoretical side, we study the impact of different $w(z)$ by using six DE parametrization models,
including the CPL, the JBP, the BA, the MZ, the FSLL and the WANG model.
In addition, we also discuss the effects of spatial curvature on this topic.
For the observational side,
we investigate the effects of three kinds of SNe Ia data, two kinds of BAO data, and four kinds of CMB data, respectively.
The detailed information of these observational data are listed in table~\ref{tab:dataset}.

Our conclusions are as follows:
\begin{itemize}

\item
(1) The evolutionary behavior of CA are insensitive to the specific form of $w(z)$
(see Table~\ref{tab:res_betaz_cmb1_bao1}, Fig.~\ref{fig:o_wz_betaz_cmb1_bao1}, and Fig.~\ref{fig:o_qz_betaz_cmb1_bao1}).
This result is consistent with the conclusion of \cite{Magana2014}.

\item
(2) Compared with the
$\Lambda$CDM model, the dynamical evolution of EoS is not favored by the current
observational data (see table \ref{tab:aicbic}). This means that, due to low statistical significance, the slowing
down of CA is still a theoretical possibility that cannot confirmed by the
current observations.

\item
(3) Considering spatial curvature or not will significantly change the evolutionary trajectories of CA:
in the framework of dynamical DE models,
a flat Universe favors an an eternal CA, while a non-flat Universe prefers a slowing down CA
(see Table~\ref{tab:res_curv_compare}, Fig.~\ref{fig:ocpl_eos_flatcompare} and Fig.~\ref{fig:o_qz_flatcompare}).

\item
(4) The use of SNe Ia data has significant impacts on CA:
for the CPL model,
SNLS3 datasets favor a slowing down CA at 1$\sigma$ CL, while JLA samples prefer an eternal CA
(see Fig.~\ref{fig:ocpl_eos_sncompare} and Fig.~\ref{fig:o_qz_sncompare}).
These results verify the conclusion of \cite{Magana2014}.

\item
(5) The effects of different BAO data on the evolutionary behavior of CA are negligible
(see Fig.~\ref{fig:ocpl_eos_baocompare} and Fig.~\ref{fig:o_qz_baocompare}).

\item
(6) Compared with CMB distance prior data, full CMB data more favor a slowing down CA in the framework of the CPL model
(see Fig~\ref{fig:ocpl_eos_cmbcompare} and Fig.~\ref{fig:o_qz_cmbcompare}).
As far as we know, the effects of different CMB data on CA have not been studied in the past.

\end{itemize}

It is clear that the evolutionary behavior of CA depends on both the theoretical models and the observational data:
for the theoretical side, it will be significantly changed by the introduction of spatial curvature;
for the observational side, it is very sensitive to the SN Ia and the CMB data.

Since SNe Ia data has significant impacts on CA,
it is very important to take into account the systematic error of SNe Ia data seriously.
In the present work, only the evolution of SN color-luminosity parameter $\beta$ is considered.
Other factors, such as the evolution of intrinsic scatter $\sigma_{\rm int}$~\citep{Marrinerl2011},
the different choice of SN lightcurve fitter model~\citep{Bengochea2014,HLLW2015a,HLLW2015b},
and the different method of calculating SN lightcurve parameters~\citep{Dai2015},
can also affect the cosmology-fit results given by SNe Ia data.
These topics deserve further investigations in future.

One of the keys of exploring CA
is to break the degeneracy between the spatial curvature and the EoS parameters with more data.
It will be interesting to study the impacts of other cosmological observations
(such as the Hubble parameter, the weak lensing, and the growth factor data) on CA.
This issue will be studied in a future work.

\begin{acknowledgements}

ML is supported by the National Natural Science Foundation of China (Grant No. 11275247, and Grant No. 11335012)
and 985 grant at Sun Yat-Sen University.
SW is supported by the National Natural Science Foundation of China under Grant No. 11405024
and the Fundamental Research Funds for the Central Universities under Grant No. N130305007.
\end{acknowledgements}


\end{document}